\documentclass[12pt,preprint]{aastex}

\def\tighttable{\def\baselinestretch{1.0}}
\def\arcsec{\ifmmode '' \else $''$\fi}
\def\arcmin{\ifmmode ' \else $'$\fi}
\def\arcsecpoint{\ifmmode ''\!. \else $''\!.$\fi}
\def\arcminpoint{\ifmmode '\!. \else $'\!.$\fi}
\def\Hubble{\ifmmode {\rm km\,s}^{-1}\,{\rm Mpc}^{-1}
	\else km\,s$^{-1}$\,Mpc$^{-1}$\fi}
\def\ergsec{\ifmmode {\rm ergs\;s}^{-1} \else ergs s$^{-1}$\fi}
\def\ergscm{\ifmmode {\rm ergs\,s}^{-1}\,{\rm cm}^{-2}
	  \else ergs\,s$^{-1}$\,cm$^{-2}$\fi}
\def\ergscmA{\ifmmode {\rm ergs\,s}^{-1}\,{\rm cm}^{-2}\,{\rm \AA}^{-1}
	  \else ergs\,s$^{-1}$\,cm$^{-2}$\,\AA$^{-1}$\fi}
\def\ergscmHz{\ifmmode {\rm ergs\,s}^{-1}\,{\rm cm}^{-2}\,{\rm Hz}^{-1}
	  \else ergs\,s$^{-1}$\,cm$^{-2}$\,Hz$^{-1}$\fi}
\def\Msun{\ifmmode M_{\odot} \else $M_{\odot}$\fi}
\def\Lsun{\ifmmode L_{\odot} \else $L_{\odot}$\fi}
\def\qo{\ifmmode q_{0} \else $q_{0}$\fi}
\def\Ho{\ifmmode H_{0} \else $H_{0}$\fi}
\def\cc{\ifmmode {\rm cm}^{-3} \else cm$^{-3}$\fi}
\def\cl{\ifmmode {\rm cm}^{-2} \else cm$^{-2}$\fi}
\def\micron{\ifmmode \mu{\rm m} \else $\mu$m\fi}
\def\kms{\ifmmode {\rm km\,s}^{-1} \else km\,s$^{-1}$\fi}
\def\ltsim{\raisebox{-.5 ex}{$\;\stackrel{<}{\sim}\;$}}

\newcommand {\HST}{{\it HST}}
\newcommand {\FUSE}{{\it FUSE}}
\newcommand {\VLT}{{\it VLT}}
\newcommand {\lya}{Ly$\alpha$}
\newcommand {\lyb}{Ly$\beta$}
\newcommand {\lyc}{Ly$\gamma$}

\newcommand {\he}{\ion{He}{2}}
\newcommand {\h}{\ion{H}{1}}
\newcommand {\lm}{$\lambda$}

\slugcomment{To Appear in the {\it Astrophysical Journal}}
\shorttitle{Intergalactic helium}
\shortauthors{Zheng et al.}

\received{2003 July 31}
\begin{document}

\title{A Study of the Reionization History of Intergalactic Helium\\
with {\it FUSE} and {\it VLT}\altaffilmark{1}}

\author{
W. Zheng\altaffilmark{2}, 
G. A. Kriss\altaffilmark{2,3},
J.-M. Deharveng\altaffilmark{4},
W. V. Dixon\altaffilmark{2},
J. W. Kruk\altaffilmark{2},
J. M. Shull\altaffilmark{5,6},
M. L. Giroux\altaffilmark{7},
D. C. Morton\altaffilmark{8},
G. M. Williger\altaffilmark{2},
S. D. Friedman\altaffilmark{3},
and H. W. Moos\altaffilmark{2}
}

\altaffiltext{1}{Based on 
observations made for the Guaranteed Time Team
by the NASA-CNES-CSA \FUSE\ mission, operated by the Johns Hopkins University
under NASA contract NAS5-32985;
Ultraviolet-Visual Echelle Spectrograph 
observations performed at the European Southern Observatory, Paranal, Chile, 
within the program 68.A-0230; and observations with the NASA/ESA {\it Hubble 
Space Telescope}, obtained at the
Space Telescope Science Institute, which is operated by the Association of 
Universities of Research in Astronomy, Inc., under NASA contract NAS5-26555.}
\altaffiltext{2}{Center for Astrophysical Sciences, Department of Physics and
Astronomy, The Johns Hopkins University, Baltimore, MD 21218
(zheng@pha.jhu.edu; wvd@pha.jhu.edu; kruk@pha.jhu.edu; williger@pha.jhu.edu;
hwm@pha.jhu.edu)}
\altaffiltext{3}{Space Telescope Science Institute, 3700
 San Martin Drive, Baltimore, MD 21218 (gak@stsci.edu; friedman@stsci.edu)}
\altaffiltext{4}{Laboratoire d'Astrophysique de Marseille, BP 8,
   13376 Marseille Cedex 12, France (jean-michel.deharveng@oamp.fr)}
\altaffiltext{5}{CASA, Department of Astrophysical and Planetary Sciences,
   University of Colorado, Boulder, CO 80309 (mshull@casa.colorado.edu)}
\altaffiltext{6}{Also at JILA, University of Colorado and National Institute
   of Standards and Technology}
\altaffiltext{7}{Department of Physics and Astronomy, Box 70652, East Tennessee
 State University, Johnson City, TN 37614 (giroux@polar.etsu.edu)}
\altaffiltext{8}{Herzberg Institute of Astrophysics, National Research Council,
 Victoria, BC V9E 2E7, Canada (don.morton@nrc.gc.ca)}

\begin{abstract}
We obtained high-resolution {\it FUSE} ($R \sim$ 20,000) 
and {\it VLT} ($R \sim$ 45,000) spectra of the quasar HE2347--4342
to study the properties of the intergalactic medium
between redshifts $z = 2.0 - 2.9$.
The high-quality optical spectrum allows us to identify approximately 850 
\h\ absorption lines with column densities between
$N \sim 5 \times 10^{11}$ and $ 10^{18}\ {\rm cm^{-2}}$.
The reprocessed {\it FUSE} spectrum extends the wavelength coverage of the 
\ion{He}{2} absorption down to an observed wavelength of 920 \AA.
Source flux is detected to restframe wavelengths as short as $\sim 237 $ \AA. 
Approximately 1400 \he\ absorption lines are identified, including 917 
\ion{He}{2} Ly$\alpha$ systems and some of their \ion{He}{2}
Ly$\beta$, Ly$\gamma$, and Ly$\delta$ counterparts.
The ionization structure of \he\ is complex, with approximately 90
absorption lines that are not detected in the hydrogen spectrum. These features
may represent the effect of soft ionizing sources. The ratio
$\eta$=N(\ion{He}{2})/N(\ion{H}{1}) varies approximately from 
unity to more than a thousand, with a median value of 62 and a distribution 
consistent with the intrinsic spectral indices of quasars. This provides 
evidence that the dominant ionizing field is from the accumulated quasar
radiation, with contributions from other soft sources such as star-forming
regions and obscured AGN, which do not ionize helium.
We find an evolution in $\eta$ toward smaller values at lower
redshift, with the gradual disappearance of soft components.
At redshifts $z > 2.7$, the large but finite increase in the \ion{He}{2} 
opacity, $\tau =5 \pm 1$, suggests that we are viewing the end stages of a 
reionization process that began at an earlier epoch.
Fits of the absorption profiles of unblended lines indicate 
comparable velocities between hydrogen and $\rm He^+$ ions. For line widths
$b_{He^+}=\xi b_{H}$, we find $\xi = 0.95 \pm 0.12$, indicating a velocity 
field in the intergalactic medium dominated by turbulence.
At hydrogen column densities $N < 3 \times 10^{12}\ \rm cm^{-2}$ the number of 
forest lines shows a significant deficit relative to a power law, and 
becomes negligible below $N = 10^{11}\ \rm cm^{-2}$.
\end{abstract}

\keywords{dark matter --- intergalactic medium --- quasars: absorption lines --- quasars: individual (HE2347--4342) --- ultraviolet: general}

\section{INTRODUCTION}

According to current models of cold-dark-matter structure formation, 
the first generation of massive baryonic objects, i.e., stars, galaxies and 
quasars, was formed in the early 
universe as the result of density fluctuations of the intergalactic medium 
\citep[IGM; ][]{zhang,me96,bi}. The radiation from these objects gradually 
ionized the surrounding IGM, ending the so-called ``dark age.''
Most models predict that the intergalactic hydrogen was ionized at 
$z \sim 7-10$, and 
helium at $z \sim 3-4$ \citep[ and references therein]{barkana,loeb}.
The recent {\it WMAP} results \citep{map} provide evidence that the IGM 
reionization may have started at $z \sim 17$. \citet{cen} suggests that
there may have 
been a second stage of hydrogen reionization that took place at $z\sim 6$,
and \citet{venkatesan03} suggest that helium may also have been reionized 
twice, partially by an early generation of Population {\sc iii} stars, and 
subsequently by quasar radiation at $z < 5$.

Our knowledge of the IGM has been derived largely from the \lya\ 
forest absorption of neutral hydrogen. The expansion of the universe following 
reionization leads to a rapid drop in 
the hydrogen \lya\ effective optical depth of the IGM, $\tau \propto
(1+z)^{3.5}$ \citep{prs93}. The number density
of forest lines also increases toward lower column density, with $dn/dN \propto
N^{-1.5}$ \citep{kim}. Lines at a column
density of $N \sim 10^{13}$ \cl\ represent the mean density the IGM. The
underdense regions, represented by hydrogen column densities below $N \sim 
10^{12}$ \cl, have been largely undetected.  Throughout this paper,
the column density $N$ refers to hydrogen, unless noted otherwise.
 
The IGM can also
be traced by \he\ absorption, which is far more sensitive than \h. 
Under the hard photoionizing conditions at $z \sim 3$, He$^+$ outnumbers H$^0$ 
by a factor $\eta$ of the order of 100 \citep{dkz,gak}.
The \he\ optical depth is therefore
$\eta/4$ or approximately 25 times greater than that of \h\ for turbulent 
line broadening and $\eta/2$ for thermal broadening. For a tenuous 
structure that produces an \h\ optical depth less than 0.01, even the best 
optical spectra may not provide a significant detection. But the
corresponding \he\ optical depth, at 0.25, can be detected at a moderate 
signal-to-noise (S/N) ratio. \he\ absorption is
therefore an effective tracer of the low-density structures that fill
$\sim 70\%$ of the volume of the universe at $z \sim 3$.
Significant \he\ absorption exists even in the spectral voids where there are 
no detected \lya\ forest lines. As \citet{croft} show, each of several
different CDM models can explain the distribution of the \h\ forest lines, 
but the various models predict different results for \he\ absorption. 

The search for the \he\ Gunn-Peterson effect has yielded significant results
in recent years. \citet{jakobsen94} reported a sharp flux cutoff at
the \he\ \lya\ wavelength 304(1+z) \AA\ in the spectrum of quasar Q0302-003 
($z=3.286$), obtained with \HST.  Follow-up observations at a
higher S/N level and spectral resolution \citep{hogan,heap} 
found an average \he\ optical depth $\tau = 4.0
\pm 0.5$ at $z \sim 3$.  At far-UV wavelengths below 1200 \AA, 
\he\ spectra exhibit a significantly lower \he\ opacity, enabling detailed 
studies of the ionization structure of the IGM.  
Hopkins Ultraviolet Telescope (HUT) observations of 
the luminous quasar HS1700+64 \citep[$z=2.73$, ][]{dkz} revealed an
average \he\ optical depth of $1.00 \pm 0.07$, which suggests that the dominant
ionization source at $z\sim 2.5$ is from quasars instead of stars, but the low 
resolution of HUT does not allow a study of the individual absorption 
components. 

The bright quasar HE2347--4342 \citep[$z=2.885$, $V=16.1$; ][]{reimers97} is 
an extremely rare
object that exhibits a clear line of sight at restframe wavelengths around
300~\AA. Its UV flux, at a level of $2 \times 10^{-15}$
\ergscmA, is the highest among the five quasars with known \he\ \lya\
features \citep{jakobsen94,dkz,reimers97,anderson,zheng04}, enabling a 
high-resolution study of \he\ absorption with \FUSE.
Below a redshift of 2.7, \citet{gak}
resolved the \he\ \lya\ absorption as a discrete forest of absorption lines.
The column-density ratio $\eta$=N(\ion{He}{2})/N(\ion{H}{1})
ranges from $\sim 1$ to $>1000$, suggesting a mix of ionizing sources of
radiation consisting of quasars and star-forming galaxies.
However, approximately 50\% of these lines do not have detectable \h\ 
counterparts in the {\it Keck} spectrum of this quasar.
At higher redshifts, the \he\ absorption has a patchy
structure, including regions with low \he\
opacity (``voids'') and regions with high opacity and no detectable flux
(blacked-out ``troughs'').
Using high S/N \HST/STIS observations of this region,
\cite{smette} found that the $\eta$ value is generally high,
but shows large variations, from $\sim 30$ in 
the opacity gaps to a lower limit of 2300 at $z \sim 2.86$ in a region that 
shows an extremely low \h\ opacity over a 6.5 \AA\ spectral range. 
In order to explain the shape of the opacity gaps (absence of large individual 
absorption lines) at this and other wavelengths, they suggested that six 
bright soft ionization sources lie near the line of sight, for which the ratio 
between the numbers of \h- to \he-ionizing photons reaching the IGM is large.

This change in character of the \he\ absorption is similar to the large
increases in Lyman series \h\ opacity at $z>6$ seen in several quasars 
discovered by the Sloan Digital Sky Survey \citep{becker,fan,white03}.
As these authors have suggested that reionization of the \h\ in the IGM
occurred just prior to this epoch, it is likely that reionization of
$\rm He^+$ occurred slightly above a redshift of 3. Theoretical models 
\citep{madau,fardal,madau2,barkana,loeb} also indicate that the reionization
of intergalactic helium took place later than hydrogen as the number of 
\he-ionizing  photons is significantly smaller than that for hydrogen. 
These initial regions of \he\ ionization, however, would have recombined, and
only reached full ionization at lower redshift during the quasar era.
Indirect evidence based on an increase in \h\ line widths
\citep{ricotti00,theuns}
and a decrease in \h\ opacity \citep{bernardi} support the assumption that the
final stage of \he\ reionization may have taken place at $z \sim 3.2-3.4$. 

In this paper, we present results based on a reprocessed \FUSE\ spectrum of
HE2347--4342 and a new \VLT/UVES echelle spectrum.
Our previous paper \citep{gak} used only the \FUSE\ data from the LiF channels,
and a modest S/N {\it Keck} echelle spectrum between 3800 and 6000
\AA. The improved \FUSE\ pipeline allows us to extract spectra from the SiC 
channels, which extend the wavelength coverage to the 912--1000 \AA\ region.
The expanded wavelength coverage enables us to study the evolution of 
intergalactic helium with redshift down to $z = 2$,
and also to observe the Ly$\beta$ portion of the \he\ forest.
Our new \VLT\ spectrum spans the full redshift range covered by the \FUSE\ 
data, and its higher S/N makes it possible to identify weak \h\ lines with
\he\ counterparts that are not detectable in the {\it Keck} spectrum.
We follow the traditional approach in this paper of fitting individual spectral
features in the \FUSE\ and \VLT\ spectra.
In a separate paper, \citet{shull03} describe an independent analysis 
of the \FUSE\ spectrum at redshift 2.3-2.9 and 
another set of \VLT/UVES data using a non-parametric approach to study the 
opacity variations in the IGM. They also identify fluctuations in the ionizing 
radiation field on a fine-grained scale, $\Delta z \sim 0.001$.

Our discussion in this paper is organized as follows.
In \S2 we describe our reprocessing of the \FUSE\ data, and our new \VLT/UVES
observations.
In \S3 we explain how we identify and fit individual spectral features in the
\FUSE\ and \VLT\ spectra.
In \S4 we discuss the scientific results of our analysis, concentrating on
the column-density distribution of \lya\ absorbers (\S4.1),
the velocity field in the IGM (\S4.2),
the evolution of the ionization state of the IGM (\S4.3),
the evolution of the \he\ opacity (\S4.4),
and the approach to the \he\ reionization epoch (\S4.5).
Section 5 summarizes our results.

\section{DATA}

\subsection{Ultraviolet Data}

The {\it Far-Ultraviolet Spectroscopic Explorer} (\FUSE) is a NASA-CNES-CSA 
mission dedicated to high-resolution spectroscopy between 905 and 1187 \AA.
It consists of two UV detectors, four Rowland-circle gratings and four 
mirrors. The LiF-coated optics yield a wavelength coverage of 
$\sim 979 - 1187$ \AA, and  the SiC-coated optics cover 
$\sim 905 - 1104$ \AA. For a full description of \FUSE, its mission, and its 
in-flight performance, see \citet{moos} and \citet{sahnow}.

The \FUSE\ observations of HE2347--4342 were carried out in three periods in
2000:  June 27, August 18-25, and October 15-20. A total of 243 raw images
were obtained, representing an exposure time of 619 kseconds, of which 431
kseconds were at night. 
While the \FUSE\ detector background is intrinsically low, the source count
level from HE2347--4342 is still weaker than the background because of the high
spectral resolution. The key step in our data analysis is the background
subtraction. The \FUSE\ background consists of two components: the detector
dark count and the scattered light. The dark count is roughly constant
across the detector but varies in intensity with time. The scattered light 
during day time is an order of magnitude stronger than at night, and it 
produces considerable and different spatial structures across the detectors. 
The inclusion of the day portion would completely obscure the weak 
source structure. We therefore use only the night portion of the data.

Based on the data of this quasar from the LiF channels of \FUSE\ and
a {\it Keck} optical spectrum, \citep{gak} have reported
a rich absorption structure that reflects a broad range of ionization level.
The average value of $\eta \sim 80$ is consistent with photoionization of the 
absorbing gas by a hard ionizing spectrum resulting from the integrated light 
of quasars. The presence of \he\ absorbers with no \h\ counterparts indicates
that structure is present even in low-density regions.

A new version of the \FUSE\ calibration software (CalFUSE v2.2) enables us
to reprocess the data and improve its quality.
We first process each of the 243 raw exposures, rejecting undesirable events
such as detector count bursts.
The processed spectra are then merged to form a final image with the highest
S/N possible. From this image, we are able to model and subtract the
background more accurately than before.

We perform the background subtraction using two parallel approaches,
including the standard pipeline procedure and manual extraction. 
To extract the spectrum manually, we define respective background windows
on both sides of the source extraction windows on the detector. The background 
variations are assumed to be linear across the spatial dimension (Y), and its
slope is estimated from the two background windows. The background level at
the extraction window is then interpolated at every wavelength unit along the 
dispersion direction (X).
This method cannot be fully implemented for extraction windows 
near the detector edge. In such cases, we define two background windows
on one side of the extraction window, calculate the slope of variations and 
extrapolate the background level at the source position. A pair of background 
windows is necessary because the background varis across the detector,
even at a low level.
 
The manual extraction yields 
better results, as judged by the level of residual flux in the opaque portions
of the \he\ absorption spectrum.
The standard pipeline extraction, on the other hand, produces a better
photometric agreement at long wavelengths in the region of overlap with the
HST/STIS data.
The final data we present here are based on the manual extraction using
photometric corrections as described below.

For a given window
position, we extract the data using different sets of detector 
pulse-height parameters \citep{pulse}: 4-16, 5-16, 7-15, and 8-12. 
In principle, dark current and photon events exhibit different
pulse characteristics. By choosing a narrow band of the pulse height, one
would be able to reduce the contribution of dark-current events. However, a
fraction of the source counts would also be lost, leading to photometric 
inaccuracy. The standard parameters for the pulse height, between 4 and 16 
counts, yield a flux level that is close to photometric. 
A pulse height range of 5--16, however, maximizes the S/N, so
we choose this range and normalize the extracted spectrum to the same
flux level as obtained using the photometric pulse height range of 4--16.

The full set of \FUSE\ data consists of eight segments labeled LiF 1A, 1B, 2A,
2B and SiC 1A, 1B, 2A, 2B. 
Sky emission lines are removed from the spectra \citep{ag}. 
Significant airglow lines include \ion{He}{1} \lm 584 (second order at 1168 
\AA), \ion{N}{1} \lm 1134, \ion{O}{1} \lm 989 and the Lyman-series lines. 
The spectra are then binned to 0.025 \AA, and combined with their S/N as the
weighting factor at every pixel.
This level of binning gives critical sampling of two bins per $R=20,000$
resolution element provided by \FUSE, and corresponds identically in velocity
($\sim 7.5$ \kms) to the 0.1 \AA\ bins we use in sampling
the \h\ spectrum described later in \S2.2.
A typical S/N level per resolution element is 10 at the long
wavelength part ($\lambda > 1050$ \AA), and 3 at the short wavelengths.

The spectral normalization involves two steps: First, we correct the zero
level. While the spectra extracted with CalFUSE v2.2 improve upon
earlier pipeline results, and our manual extraction technique is even better,
there are still regions where the zero level is not perfect.
This is particularly obvious in the SiC
spectra from detector 1, where the SiC 1A and 1B extraction windows lie 
near the edge of the detector. 
We correct the zero level using the \h\ \lya\ forest spectrum where
there are many features with fully saturated absorption at the line center.
It is reasonable to assume that their
\he\ counterparts will also be completely absorbed at the line center, as
the \he\ optical depth is considerably higher than that of the \h\ feature. 
We assume that these data points should have zero flux, and fit
a smoothly varying polynomial to these low points to determine the zero-level 
correction.  This correction is up to 10\% of the maximum flux in the LiF data 
segments, and up to 40\% in the SiC channels.

Secondly, we normalize the \FUSE\ spectrum, assuming the intrinsic, unabsorbed 
spectrum is a reddened power law. Since more than a thousand \he\ absorption 
lines nearly blanket the entire \FUSE\ spectrum, it is difficult 
to determine the continuum level from the \FUSE\ spectrum itself. As described 
by \citet{gak}, a pair of \HST/STIS spectra are used to 
derive the intrinsic continuum.
These data were obtained contemporaneously with the respective \FUSE\ data
and serve as a photometric standard for our extracted \FUSE\ spectrum.
A comparison between the STIS and \FUSE\ spectra indicates that the photometric
quality of the \FUSE\ data is reasonably good, agreeing with the STIS data
to within about 5\% between 1150 and 1187 \AA.
The best-fit power law is 
$f_\lambda = 3.3 \times 10^{-15}$($\lambda$/1000\AA)$^{-2.4}$
\ergscmA, with E(B-V)=0.014. 
An extrapolation of this continuum model serves as the baseline for the 
renormalization of our \FUSE\ spectrum.

\subsection{Optical Data}

Optical echelle spectra of HE2347--4342 were obtained at the ESO \VLT/UVES on
2001 November 23-24. The exposure time for each readout was 30 minutes, and the
total exposure time was 8 hours. The sky was photometric. With
a slit opening of 1 arcsecond, the spectral resolution is approximately 45,000.
We used the blue instrument arm, consisting of a $2048 \times 4096$ CCD
detector, an echelle grating of 41.59 line mm$^{-1}$, and a cross disperser of 
600 line mm$^{-1}$ and blaze wavelength of 4600 \AA. With a central wavelength 
4200 \AA\ for the cross disperser, the spectrum covers the wavelength range 
between 3600 and 4800 \AA\ in 31 orders, which corresponds to the entire
\he\ \lya\ wavelength range studied with \FUSE\ from $z = 2.0$ to 2.9.
There are significant overlaps in wavelength coverage between adjacent orders.
The wavelength calibration spectra were taken with the ThAr lamp.

Using IRAF tasks designed for echelle data, we processed the spectra.
For each extraction segment, we used 10-15 pixels on both sides of the trace
to derive the background.
The final wavelength calibration is accurate to $\sim 0.007$ \AA, and it has
been corrected to vacuum wavelengths in a heliocentric reference frame.
We normalized each individual echelle order with a polynomial, then rebinned 
them to 0.1 \AA. We then merged all 31 segments of normalized spectra,
weighted by the individual S/N ratios.
The S/N ratio in the merged spectrum is about 110 per 0.1 \AA\ bin
at 4700 \AA, and about 46 at 3850 \AA. This is approximately 2.5 and 10 
times that of the {\it Keck} data, respectively. Only the \VLT\ data cover
the wavelength range between 3600 and 3800 \AA. To improve the continuum 
fitting, this count spectrum is further normalized with a high-order 
polynomial across the entire wavelength range. 
Figure 1 displays the normalized \FUSE\ and \VLT\ spectra of HE2347--4342
across the full wavelength range covered by our new data.

\section{LINE FITTING}

We analyze the spectral features in the \FUSE\ and \VLT\ spectra using
the IRAF task {\it Specfit} 
\citep{specfit}. Absorption features are fit with Gaussian
profiles rather than Voigt profiles. This is
a good approximation for line components with column densities below $10^{15}$ 
\cl.  Saturated lines may be better represented by Voigt profiles, but
it is not important for our purpose since our main goal is to study the weak 
absorption lines. Because of the large number of components and 
variables, we carry out the fitting procedure in 48 wavelength segments of 
approximately 25 \AA\ each.

Below 3984 \AA\ in the optical spectrum, \lyb\ absorption starts to appear. 
Line fitting in this region is carried out using both
\lya\ and \lyb\ for each identified feature. Paired absorption lines
are fit using the same velocity width and column density, and their 
central wavelengths are tied to the ratio of their intrinsic wavelengths.

Free parameters in the fit are line width, column density, and the centroid
wavelength. The weakest features that we are able to identify in the \VLT\ 
spectrum have a column density $N = 5 \times 10^{11}$ \cl, mostly at 
wavelengths longward of 4400 \AA. At the short-wavelength
end, the minimum detectable column density is $\sim 1.5 \times 10^{12}$ \cl.
The number of these weak lines and their fitted parameters are affected
by the continuum level. The errors in their column densities may
be as large as  a factor of two.
We identify metal lines as those with $b$ values less than 10 \kms.
A total of 852 \lya\ absorption lines are present in the \VLT\ spectrum over 
the full 3600-4750 \AA\ wavelength range. 

The initial line fits to the \FUSE\ spectrum are based on our results from 
fitting the \h\ \lya\ forest. For every \h\ absorption line, we define a \he\ 
counterpart with the same velocity width (see \S 4.2) and redshift, with only 
the column density 
allowed to vary. As described in \citet{gak}, additional \he\ absorption lines
are added in regions where the \he\ optical depth is so high high that no 
known \h\ component can provide with reasonable $\eta$ values. Such added 
absorption lines are given a fixed Doppler parameter $b=27$ \kms. 
For the \VLT\ spectrum, multiple Gaussian profiles are often 
introduced to fit a single absorption feature. Fitting to the corresponding 
\FUSE\ data sometimes to converge to one of these profiles, but fails to 
identify the others. In such a case, we fix this set of profiles to the same
$\eta$ value, as these absorbers are ionized by the same radiation field.
Approximately ten \h\ absorption lines have no apparent \he\ counterparts, 
and they are not apparently blended. We assign \he\ column densities to 
these lines as 1-$\sigma$ lower limits that are on the order of $10^{13}$ \cl,
and mark them in Table 1 with null errors in the \he\ column density. Most of 
these absorption lines 
are only marginally detected in the \VLT\ data; therefore their reality may be 
questionable. It is worth noting that uncertainties in the \FUSE\ data 
reduction and fitting, particularly the imperfect background subtraction may 
contribute significant errors in these peculiar absorption lines.

At wavelengths shortward of 996 \AA, the presence of \he\ \lyb\ poses 
additional difficulty in line identification.
We fit the \FUSE\ data in this region in two ways.
For lines that are not badly blended, we fit \he\ \lya\ and \lyb\ 
simultaneously, using the same approach we applied to the \VLT\ \h\ spectrum.
The \lya\ and \lyb\ lines are forced to have the same column density, redshift,
and Doppler parameter, 
For those lines that are blended, the noisy \lyb\ region can adversely affect
the fitting process. In such cases, we determine the \he\ column densities of
the \lyb\ lines from the longer-wavelength \lya\ lines alone and fix this
parameter for the corresponding \lyb\ line.
We then fit the \he\ \lya\ lines in the blended region using the corresponding
\h\ lines from the \VLT\ spectrum to fix the redshifts and line widths.
As above, if additional \he\ opacity is needed at specific wavelengths,
we add additional \he\ \lya\ components with a fixed Doppler parameter.
We do not fit the spectrum below 920 \AA, as blended
Lyman-series airglow contamination becomes too strong to correct.

At wavelengths longward of 1000 \AA, our fits agree well with those in 
\citet{gak}, but require fewer added \he\ components. 
The higher quality of our \VLT\ spectrum 
allows us to identify more weak hydrogen absorbers. Between 1000 and 1130 
\AA, we identify 362 absorption lines with both \h\ and \he\ \lya\ absorption
and only require 43 additional \he\ \lya\ lines with no \h\ counterpart.
The number of 
identified \h\ \lya\ forest lines is nearly twice that in the {\it Keck} 
spectrum of this object \citep{songaila,gak} over this spectral range,
and they include most of the added \he\ \lya\ components in the work of 
\citet{gak}. 
The properties of sample components are listed in Table 1: their absorption
redshifts, \he\ column densities, Doppler parameters, 
\h\ column densities, and $\eta$ values. Null errors in the Doppler 
parameter represent an added component that is not detected in the \VLT\ data
(the corresponding \h\ column density is the upper limit.) In such a case, the
$\eta$ value is an upper limit.

\section{DISCUSSION}

\subsection{Distribution of Absorption Features}

The \h\ \lya\ forest lines in quasar spectra follow a well-defined power-law 
distribution, $dn/dN \propto N^{-\beta}$, where $\beta =1.5$ at $3 \times 
10^{12} < 
N < 10^{15}$ \cl\ \citep{tytler,rauch,kim}. At higher column 
density there are more damped systems than this power law predicts.
Below $N < 3 \times 10^{12}$ \cl, line identification becomes difficult, and 
the results are not certain. \citet{hu} suggest that this power law may be 
extended to lower column density, assuming a large correction factor. 

As the \FUSE\ data reveal a large number of absorption lines with low
\h\ column density, useful information about their distribution with column 
density may be obtained. We first examine the column-density distribution of
the \lya\ lines that are observed in both \h\ and \he.
In Figure 2, the distribution of the \h\ forest lines in HE2347--4342 is 
plotted. The results are consistent with a power law of $\beta=1.5$ between
$3 \times 10^{12} < N < 10^{16}$ \cl. Another histogram displays the 
distribution of values derived from the \he\ 
column density. Each data point is divided by 62, the median $\eta$.
The distribution of these calculated hydrogen absorption lines follows a
pattern similar to the real ones, demonstrating the possibility of using \he\ 
parameters to describe the \h\ counterparts. The histograms do not match well 
at higher column density, as a result of this simplified approach which
assumes a fixed value for $\eta$.

We extend the database for Figure 2 with additional absorption lines which are 
detected only in \he. If the average value of $\eta=300$, approximately
100 data points are added to the \he\ database in Figure 2, to produce the
results displayed in Figure 3. If we choose
a value lower than 300, many absorption-lines should have been detected in the 
optical spectrum, and values higher than 300 will leave a gap in the 
distribution shown in Figure 3. The added points
represent our estimate of the low-column-density end of the \h\ absorption line
distribution. There is a significant downturn around $N = 3 \times 
10^{12}$ \cl, from 
$\beta \sim 1.5$ to $0.3$. On the basis of this pattern, we infer that
the number of absorption lines with $N < 10^{11}$ \cl\ 
is negligible. 

A part of the observed trend in Figures 2 and 3 at the low-column-density 
end may be attributable to the confusion limit of the \FUSE\ data: Weak 
absorption features fall into other troughs and become undetected. 
We have carried out simulations by introducing additional absorption lines 
into the \FUSE\ spectrum, and performing the same line-fitting procedure 
described in \S3.
At a \he\ column density of $\sim 10^{15}$ \cl, more than 90\% of the 
absorption lines are identified. At $\sim 10^{14}$ \cl, approximately 
two thirds of the lines are identified. The others are heavily blended, and 
the fitting task only yields a single fitted component for them. This 
may explain why many fitted lines have Doppler parameters greater 
than 40 \kms. At the limiting column density of $10^{13}$ \cl, only one out of
five may be detected. Assuming an average $\eta = 300$, these components have 
their \h\ column density on the order of $10^{10}$ \cl. A correction factor of
5 may therefore increase the number of weak absorbers at $N< 10^{11}$ \cl. 
The downturn of the distribution shown in Figure 3 takes place at a \he\ 
column density of $N \sim 10^{15}$ \cl, where the correction factor is
insignificant. We therefore suggest that this change in the distribution is 
real, not just the result of confusion.

\subsection{Velocity Field in the IGM}

An important but as yet unknown parameter in the IGM is the velocity ratio 
$\xi = b_{He}/b_{H}$ between the $\rm He^+$ ions and hydrogen atoms in an
individual spectral feature.
This is a vital clue to the actual physical processes that dominate the
evolution of structure in the IGM.
Previous models and observations often
assumed a value, either 1.0 for purely turbulent motions, or 0.5 for 
purely thermal broadening, in order to constrain other parameters.
Even with the high spectral resolution of \FUSE, an attempt to derive the
\he\ Doppler line widths is difficult because of line blending and the
low S/N.
In spectral voids, however, where absorption lines are sparse, we have
identified a sample of individual \he\ and \h\ absorption lines that are not
blended in either \lya\ or \lyb.
We fit the profiles of these individual absorption features
(Figures 4 and 5) and let the velocity widths vary as free parameters.
The optical depths of the \lya\ and \lyb\ lines are also linked by simple
atomic physics, providing an important additional constraint on the allowed
range of the fitted line widths.
Table 2 summarizes the results of our fits.
A linear fit to the line widths yields a ratio of the \he\ to \h\ line widths 
of $\xi = 0.95 \pm 0.12$. As shown Figure 6, this velocity ratio is consistent
with turbulence being dominant, with one or two cases of possible thermal 
line broadening. We therefore have assumed identical \he\ and \h\ line widths
for all absorption lines in our analysis.

\subsection{Evolution of Ionization Level}

The derived $\eta$ ratios are plotted in Figure 7, along with the lower limits 
from those \he\ \lya\ lines which do not have \h\ counterparts
(now a minor fraction of the data set, $<10\%$) and the upper limits from the
\h\ lines without significant \he\ counterparts.
Our analysis reveals a significant population of 
IGM structures with hydrogen column densities between $5 \times 10^{11}$ 
and $2\times 10^{12}$ \cl. They account for approximately 40\% of all the 
identified \he\ absorption lines. We calculate the average value of $\eta$
in logarithmic units, including only lines that are detected in both 
the \FUSE\ and \VLT\ spectra.
For {\it n} data points, we calculate the logarithmic average as
$\log(\bar{\eta}) = (\sum \log(\eta_i))/n$, and its corresponding uncertainty
as $\log(\overline{\Delta \eta}) = (\sum (\log(\Delta \eta_i))^2)^{1/2}/n$. 
The average value of $\eta \sim 48 \pm 10$ strongly suggests that the main
ionization source is the accumulated emission from quasars.
\citet{zheng} and \citet{telfer} derive a mean spectral index below 1000 \AA\ 
for quasars of $\alpha \sim -1.7$, which is consistent with 
the ionization level measured here. The median value of $\eta$ is 62, which is 
reasonably close to the average value. 
However, there is significant variation of the ionization parameter $\eta$,
from less than unity to over a thousand. Some lines with low
$\eta$ values in our list are associated with heavily saturated Lyman-limit 
systems, in which the measurement errors are large. 

Cases with $\eta> 500$ may represent the regions in the IGM that are ionized 
by local soft radiation sources, but not yet by the metagalactic UV background 
radiation field. Approximately 90 such lines are added in fitting the 
\FUSE\ 
spectrum, particularly at $z>2.7$. They are needed to explain the complete 
absorption troughs, and they confirm the suggestion of \citet{zheng98} that
a significant part of the \he\ absorption may be produced by absorption lines 
without detected \h\ counterparts. These lines represent
the tenuous regions of the IGM that have not been detected in hydrogen.
The lack of such high $\eta$ values at lower redshifts suggests that
the ionization of intergalactic helium is fairly complete at $z \sim 2$.

Figures 7 and 8 show that the column-density ratio 
$\eta$ decreases gradually toward lower redshift.
The main reason appears to be the decline in the number of absorption lines with
$\eta > 1000$ with decreasing redshift (only one is seen below $z = 2.25$),
as the \he\ reionization by a 
metagalactic UV background field becomes complete. \citet{madau} and 
\citet{madau2} model the evolution of 
the ionizing continuum under the assumption that
quasar emission is the dominant source of ionizing photons and suggest that
the \he/\h\ ratio in the IGM and the \lya\ clouds increases from 
$\sim 25$ at $z = 0$ to $\sim 45$ at $z = 2.5$, and then decreases again below 30 
for $z > 4.5$. \citet{telfer} study the EUV continuum shape of quasars and find 
that it remains fairly constant between redshifts of 0.3 and 2.
Such a trend is also noted in the model results of \citet{fardal} for a 
fixed continuum-power-law index.
Our results strongly suggest that the main ionizing source at redshift 
$z \sim 2.8$ is the accumulated radiation from quasars. 

As suggested by \citet{gak} and \citet{shull03}, the large variations in 
$\eta$ may be due to local sources of ionizing radiation.
For each observed value of $\eta$, we infer the incident spectral index
$\alpha_{EUV}$ (defined by $f_\nu \propto f^{-\alpha_{EUV}}$) using models of
power-law radiation propagated through the IGM \citep[ model A2
in their Figure 10]{fardal}.
In Figure~9 we compare this distribution of inferred spectral indices to
those measured by \citet{telfer}.
At low values of $\alpha_{EUV}$ the two distributions are similar, suggesting
that the observed variations in $\eta$ are consistent with the natural
variation of intrinsic quasar spectral indices.
At high values of $\alpha_{EUV}$, however, the excess of features with
high $\eta$ indicates a contribution from an additional population of
soft ionizing sources such as star-forming galaxies or obscured AGN.
One problem with this simple interpretation is that the observed space
density of quasars at $z = 2.5$ is too sparse to produce fluctuations on the
observed scale of $\Delta z = 0.001$, and the expected high space density of
star-forming galaxies should produce a more uniform radiation field.
As discussed by \citet{shull03},
the combination of the natural variation in quasar spectral indices, the 
resulting difference in the proximity spheres for \h\ and \he, and radiative 
transfer through a non-uniform IGM may all conspire to produce the observed
variations in $\eta$.

\subsection{Evolution of \he\ Opacity}

Studies of \he\ absorption along several lines of sight \citep[ and references
therein]{dkz,heap,gak} 
have shown that from $z=2.3$ to $z=3.3$ the average 
opacity increases toward higher redshift. 
In our reprocessed \FUSE\ spectrum of HE2347--4342, for a given section of the
spectrum, we calculate the average transmission and its associated uncertainty
by summing over the {\it n} data points $\overline{T} = (\sum t_i)/n$ and 
$\overline{\Delta T} = (\sum (\Delta t_i)^2)^{1/2}/n$, where the transmission
at each point $i$ in the normalized binned spectrum is given by $t_i$.
The \he\ \lya\ optical depth is then calculated as
$\tau_\alpha = -{\rm ln~\overline{T}}$.
At wavelengths shorter than 995.5 \AA, however, we must modify this procedure
as \he\ \lyb\ and other higher-order Lyman series lines begin to blend in
with the \he\ \lya\ forest.
The high resolution of the \VLT\ and \FUSE\ spectra enable us to separate the
\lya\ absorption lines at $z < 2.27$ from high-order Lyman lines at higher
redshift.  To isolate the contribution from the \lya\ absorption lines alone, 
we construct a model for the short-wavelength section of the
spectrum based on the \h\ line list from the fitted \VLT\ spectrum (for the
\he\ \lya\ lines) and the fitted parameters of the longer-wavelength
\he\ \lya\ lines (for the higher-order Lyman lines).
At wavelengths shortward of 995.5 \AA\ ($ z < 2.28$), we then calculate the
\he\ \lya\ opacity from this modeled spectrum.
Figure 10 shows the average \he\ \lya\ optical depth as a function of redshift.
For comparison, we show the empirical trend with redshift for \he\ of
$\tau \propto (1+z)^{3.5}$ \citep{fardal}.
At redshifts below 2.7, the agreement is quite reasonable;
however, at higher redshift, the opacity grows much more rapidly.
We discuss this in the next section.

\subsection{He II Opacity at $z > 2.7$}
At redshifts above $z = 2.7$, the spectrum of HE2347--4342 is largely opaque
in \he\ \lya.
The high S/N \HST/STIS spectrum \citep{smette} shows
several deep absorption troughs at $z > 2.77$.
Following the nomenclature and wavelength intervals defined in their Table 3,
Troughs A, C, D, and I have average \he\ \lya\ optical depths of $\sim 4$ or 
more.
Our new short-wavelength \FUSE\ data allow us to derive more accurate optical
depths in these deepest troughs since we can now measure the opacity not only
in \lya, but also in the corresponding higher-order Lyman lines.
For example, trough A is seen in \lya\ (1175.6-1178\AA), \lyb\ 
(991.8-993.8\AA), and \lyc\ (940.5-942.4\AA).
Because of these low S/N and uncertain continuum, the \lyc\ troughs (visible 
in Figure 1) are not quantitatively useful in improving the
limits on the \lya\ opacity.
To measure the average optical depth due to \lyb\ alone in a given
wavelength interval, we first calculate the optical depth directly from
the \FUSE\ spectrum. There are overlying \he\ \lya\ lines throughout this
region of the spectrum, however, and we correct for these by using the
predicted \he\ \lya\ opacity based on the \h\ \lya\ line list in each of
these intervals obtained from the \VLT\ spectrum. For each \h\ \lya\ line,
we assume the median value of the \he-\h\ column-density ratio $\eta = 62$,
compute the corresponding \he\ \lya\ column density, and assume that it has the
same redshift and velocity width as the \h\ feature.
We then integrate this synthetic \he\ \lya\ spectrum to obtain the 
contaminating contribution of \he\ \lya\ to the opacity in each of the regions
that we measure for the \lyb\ transmission.
Table 3 summarizes our opacity measurements for each of the wavelength
regions identified by \citet{smette}.
For each \he\ \lya\ wavelength region in column 2,
we give the redshift at the center of the region in column 3,
the \he\ \lya\ opacity measured from the \FUSE\ spectrum in column 4,
the \he\ \lya\ opacity from the STIS spectrum in column 5,
the raw \he\ \lyb\ opacity from the corresponding interval in the
short-wavelength portion of the \FUSE\ spectrum in column 6,
the corrected opacity of the \lyb\ region for overlying \he\ 
\lya\ lines in column 7,
and the inferred \he\ \lya\ opacity based on the corrected \lyb\ 
opacity in column 8.
We derive the inferred \lya\ opacity based on the higher-order Lyman lines
using the standard relationship between the optical depth and the oscillator
strengths for lines in the Lyman series:
$$
{
{\tau({\rm Ly}\alpha)}
\over
{\tau({\rm Ly}\beta)}
} = {
{f({\rm Ly}\alpha) \ \lambda({\rm Ly}\alpha)}~,
\over
{f({\rm Ly}\beta)\ \lambda({\rm Ly}\beta)}
} 
$$
\noindent where $f({\rm Ly}\alpha) = 0.4162$ and $f({\rm Ly}\beta) = 0.0791$
 are the oscillator strengths.

For the deepest troughs where only lower limits on the opacity could be set
using the STIS spectrum, we conclude that the average optical depth of \lya\ is
high, but finite, with a typical value of $\tau_{Ly\alpha} \sim 5$. 
As shown in Figure 10, the distribution of the \he\ \lya\ optical depth follows
a slowly rising empirical curve similar to that observed for the overall
\h\ opacity, and suddenly increases at $z\sim 2.85$. 
Such a sudden change in the \he\ opacity may signal the late stages of the 
reionization of intergalactic $\rm He^+$ at $z \sim 3$. An analogous phenomenon
is also observed for intergalactic hydrogen at $z \leq 6$ 
\citep{becker,fan,white03},
where the \h\ opacity abruptly rises from $\sim 3$ to $> 10$. 
The predicted increase at this epoch is more rapid than
$\tau \propto (1+z)^{3.5}$ \citep{fardal},
and the merging of ionized bubbles should produce a huge discontinuity at some
slightly higher redshift.
Numerical simulations of \h\ reionization \citep{raz02} 
illustrate this same effect.
A slow rise in opacity with redshift that roughly follows a power law
in $(1+z)$ shows first a marked departure from the power law followed by a
discontinuity of several orders of magnitude
that corresponds to the epoch when all the Str\"omgren spheres began to merge.
Our inferences from the HE2347--4342 spectrum
are consistent with indirect evidence from \h\ line widths
\citep{ricotti00,theuns} and \h\ \lya\ forest opacity
\citep{bernardi} suggesting that the \he\ reionization discontinuity occurred 
at $z = 3.2 - 3.4$, and that we are seeing only the final stages in the process
of \he\ reionization. 

\section{Summary}

We have studied absorption by \he\ and \h\ in the IGM over a redshift range
$2.0 < z < 2.9$ using deep, new \VLT/UVES
spectra and reprocessed \FUSE\ data for the quasar HE2347--4342.
The \VLT\ data reach a limiting \h\ column density of $5 \times 10^{11}$ \cl.
More than 90\% of the \he\ absorption lines identified in the \FUSE\ 
spectrum have \h\ counterparts, suggesting that 
we have traced most of the IGM mass, and that only a small fraction of it
exists in regions with \h\ column density smaller than $5 \times 10^{11}$ \cl. 
The reprocessed \FUSE\ data extend our view of \he\ absorption to wavelengths
shortward of 1000 \AA, allowing us to study its properties down to redshifts
as low as $z \sim 2$, and giving us access to the \lyb\ portion of the forest.
Using these data, we reach the following conclusions:

\begin{enumerate}

\item
The distribution function of the absorption features is a power law with 
respect to hydrogen column density. The identified and added \he\ absorption lines 
in the \FUSE\ spectrum suggest a low-column-density cutoff around an \h\ column
density of $3 \times 10^{12}$ \cl, with a gradually diminishing contribution 
from lower column density features down to $\sim 10^{11}$ \cl. 

\item
By analyzing a sample of isolated \he\ features with both \lya\ and \lyb\ 
components, we are able to compare the \he\ line widths to the widths of
their \h\ counterparts. We find that the ratio of \he\ to \h\ Doppler
parameters $\xi = b_{\rm He} / b_{\rm H} = 0.95 \pm 0.12$, indicating that
the IGM may be dominated by turbulent gas motions.

\item
We find that the ratio of \he\ to \h\ column densities $\eta$ shows a gradual
evolution with redshift, decreasing from a median value of $\sim$70 at
$z = 2.85$ to $\sim$40 at $z = 2.05$. The number of \he\ absorption features
with no \h\ counterparts, indicative of IGM regions photoionized by very soft
spectra, also decreases with decreasing redshift (Figure 7 and 8).
This trend suggests that by a redshift of 2, the ionization of intergalactic
$\rm He^+$ is mainly due to AGN.

\item
From $z = 2.0$ to 2.7, we see a slow increase in \he\ \lya\ optical depth.
At $z \sim2$, the average \he\ optical depth is about 0.5, increasing to
$\tau \sim 1$ at $z \sim 2.6$. A sudden increase in optical depth to
$\tau \sim 5$ is observed at $z\sim 2.87$. This trend in optical depth with
redshift is analogous to that observed in \h\ at higher redshift.

\item
We limit the \he\ \lya\ optical depths in the opaque regions of the UV spectrum
at $z > 2.7$ to values of $\tau_{Ly\alpha} \sim 5$ by
using the corresponding troughs in the \lyb\ 
portions of the spectrum to extend our dynamic range. 
This sudden, but finite, increase in optical depth suggests that we are seeing
the end of the \he\ reionization process that began at an earlier epoch as
indicated indirectly by the increase in \h\ line widths 
and \h\ opacity observed at $z = 3.2-3.4$.

\end{enumerate}

\acknowledgments

The successful execution of the \FUSE\ observations of this quasar is the
result of many years of collaborative work of the \FUSE\ science team. 
W.Z. would like to thank the staff at ESO Paranal observatory, especially 
Andreas Kaufer, for their professional help.

	This work is based on data obtained for the Guaranteed Time Team by
        the NASA-CNES-CSA \FUSE\ mission operated by the Johns Hopkins 
	University. Financial support to U. S. participants has been provided 
        by NASA contract NAS5-32985.
	A portion of this work is based on observations with the NASA/ESA
	{\it Hubble Space Telescope}, obtained at the Space Telescope Science
	Institute, which is operated by the Association of Universities for
	Research in Astronomy, Inc., under NASA contract NAS5-26555.

\clearpage

\centerline{\bf Figure Captions}
\bigskip

\figcaption{
Normalized spectra of quasar HE2347--4342. The upper panel shows the \FUSE\ 
data, and lower panel the \VLT/UVES data. Both the {\it FUSE} and 
VLT data are binned to 0.1 \AA. The propagated errors are plotted as dashed 
curves. 
For the \FUSE\ data, the errors are scaled down by a factor of 5 for clearer 
presentation, and are plotted only for positive
data points. Strong airglow lines are removed and marked with an Earth symbol.
Three spectral regions, where the \he\ opacity is significantly lower than 
average,
are marked as B, F, G (after Smette et al.) and K, respectively. Seven regions
of high \he\ opacity, defined by Smette et al., are also marked. 
The positions of Ly$\alpha$, Ly$\beta$ and Ly$\gamma$ are indicated at the 
quasar redshift, and the wavelengths of the interstellar Lyman-series lines 
below 950~\AA\ are labeled.
\label{fig1}}

\figcaption{
Column-density distribution of Ly$\alpha$ forest lines. The dashed line
represents the 
expected values for a power law $dn/dN \propto N^{-1.5}$. The area marked 
with ``He'' represents the values that are derived from their
\ion{He}{2} column density, divided by the median $\eta$ value of 62.
\label{fig2}}

\figcaption{
Column-density distribution of absorption systems, with added components that
are detected via only \he\ absorption in the \FUSE\ spectrum. 
To convert from \protect\ion{He}{2} column density to \protect\ion{H}{1} for these added 
components in the \FUSE\ spectrum, we assume $\eta = 300$ (marked by the scale 
at the top). Correcting for the confusion limit may increase the actual number 
of systems with column density N \protect\ltsim $10^{11}$ \cl.
\label{fig3}}

\figcaption{
Spectral features in void regions F and G. The {\it FUSE} data are binned to 
0.05 \AA. 
The arrows represent the identified absorption components. 
The Ly$\beta$ regions are blended with a number of prominent Ly$\alpha$  
absorption features at $z \sim 2.22$, some of which coincide with the 
expected Ly$\beta$ features. 
\label{fig4}}

\figcaption{
Spectral features in void region K. The {\it FUSE} data are binned to 0.05 
\AA. Some unmarked absorption features in the Ly$\beta$ regions are Ly$\alpha$ 
features at $z\sim 2.14$.
\label{fig5}}

\figcaption{
Doppler parameters for hydrogen and helium absorbers. The solid line is 
for expected values with turbulence line-broadening, and the dashed line 
for thermal broadening.
\label{fig6}}

\figcaption{
Column-density ratio $\eta$=N(\ion{He}{2})/N(\ion{H}{1}) vs. redshift. The
circles and upward arrows (colored red in the electronic edition) represent 
\ion{He}{2} Ly$\alpha$ features that have no 
H$^0$ counterparts in the {\it VLT} spectrum and thus yield only lower limits to the 
indicated $\eta$ ratios. The cross and downward arrows (colored blue in the
electronic edition) represent \ion{H}{1} Ly$\alpha$ features that have no 
\ion{He}{2} counterparts in the {\it FUSE} spectrum and thus yield only upper 
limits to the indicated $\eta$ ratios. These $\eta$ values are likely 
underestimated because of the limited \FUSE\ data quality.
At $z<2.28$, above which higher-order He~{\sc II} Lyman lines become visible,
$\eta$ values from only the Ly$\alpha$ components are shown.
The dashed curve represents the anticipated values if the ionizing sources are 
quasars with an EUV power law of $f_\nu \propto \nu^{-1.76}$, which are 
interpolated from the model results of \citet{fardal}.
\label{fig7}}

\figcaption{
Column-density ratio $\eta$ vs.\ redshift. The average $\eta$
value is calculated from the components that are detected in both the
{\it FUSE} and {\it VLT} spectra. 
The dashed curve represents the anticipated values if the ionizing sources are 
quasars with an EUV power law of $f_\nu \propto \nu^{-1.7}$, which are 
interpolated from results of \citet{fardal}.
\label{fig8}}

\figcaption{
Distribution of power-law indices of the ionizing sources. The dashed line 
represents
a scaled distribution of the $\alpha_{EUV}$ of 39 radio quiet quasars (Telfer
et al.\ 2002). 
The conversion from $\eta$ to an inferred value for $\alpha_{EUV}$ uses the 
quasar models of \citet{fardal}. The number excess at 
higher values indicates a contribution from 
soft ionizing sources such as star-forming galaxies or obscured AGN.
\label{fig9}}

\figcaption{
Redshift dependence of the \ion{He}{2} Ly$\alpha$ opacity.
Values below $z=2.3$ are derived from a line spectrum reconstructed from the
fitted parameters that omit \lyb\ and higher-order Lyman lines.
Between $z=2.3$ and 2.7 the values are direct measurements from the {\it FUSE}
data, and at $z>2.7$ they are derived from the observed \he\ Ly$\beta$
absorption region.
The curve representing $\tau \propto (1+z)^{3.5}$ is plotted for comparison.
\label{fig10}}

\clearpage

\setcounter{figure}{0}
\begin{figure}
\plotone{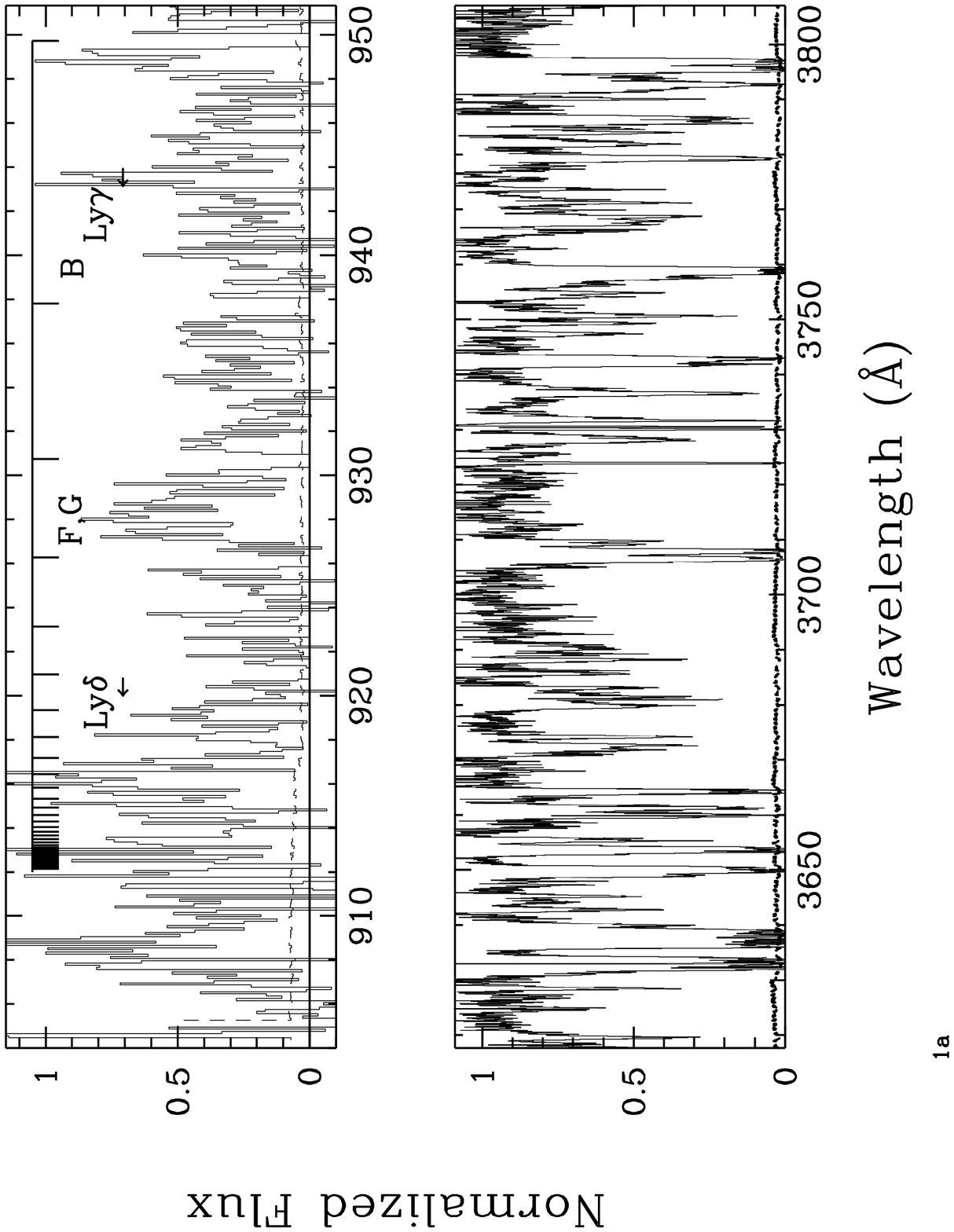}
\caption{a}
\end{figure}

\clearpage

\setcounter{figure}{0}
\begin{figure}
\plotone{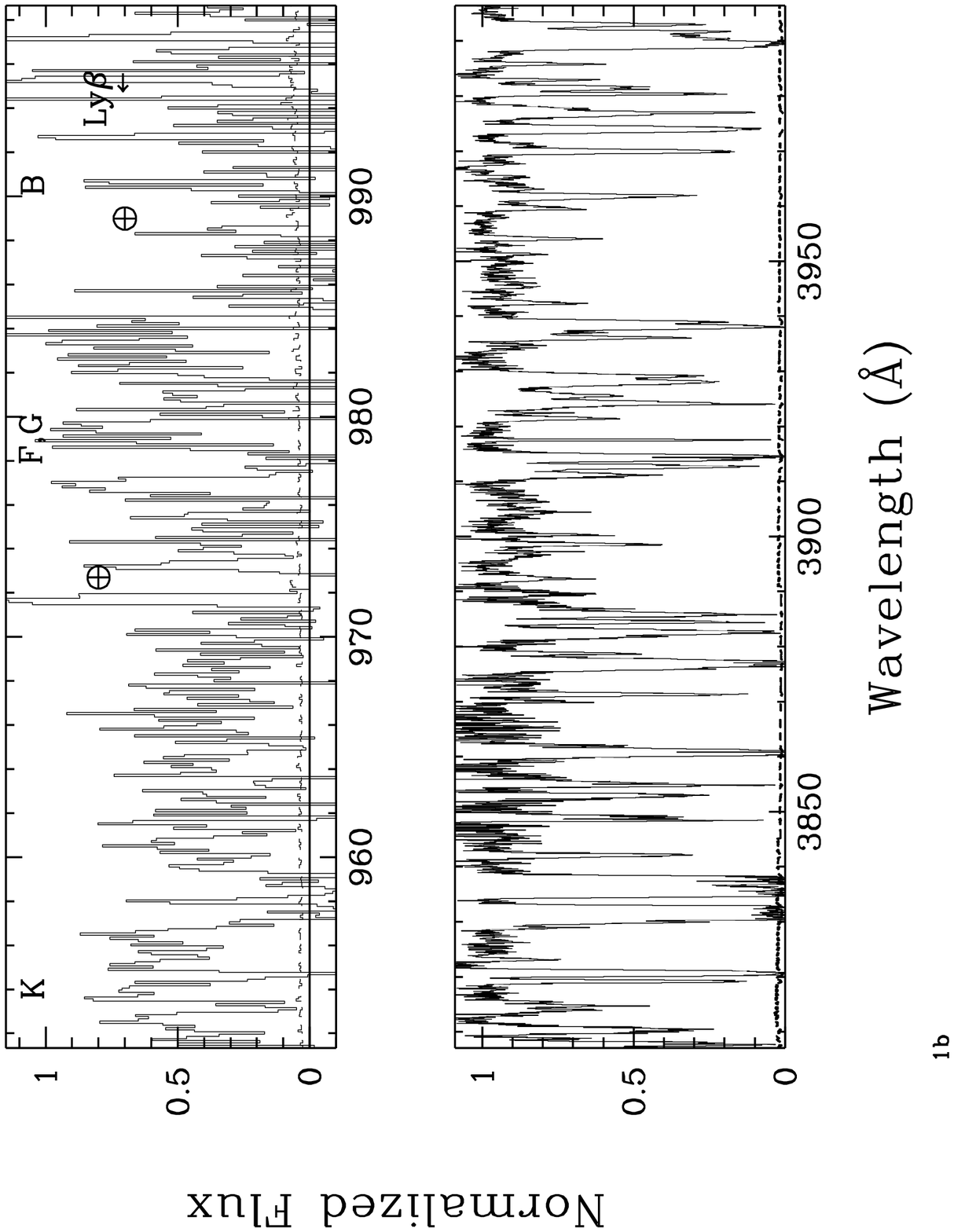}
\caption{b}
\end{figure}

\clearpage

\setcounter{figure}{0}
\begin{figure}
\plotone{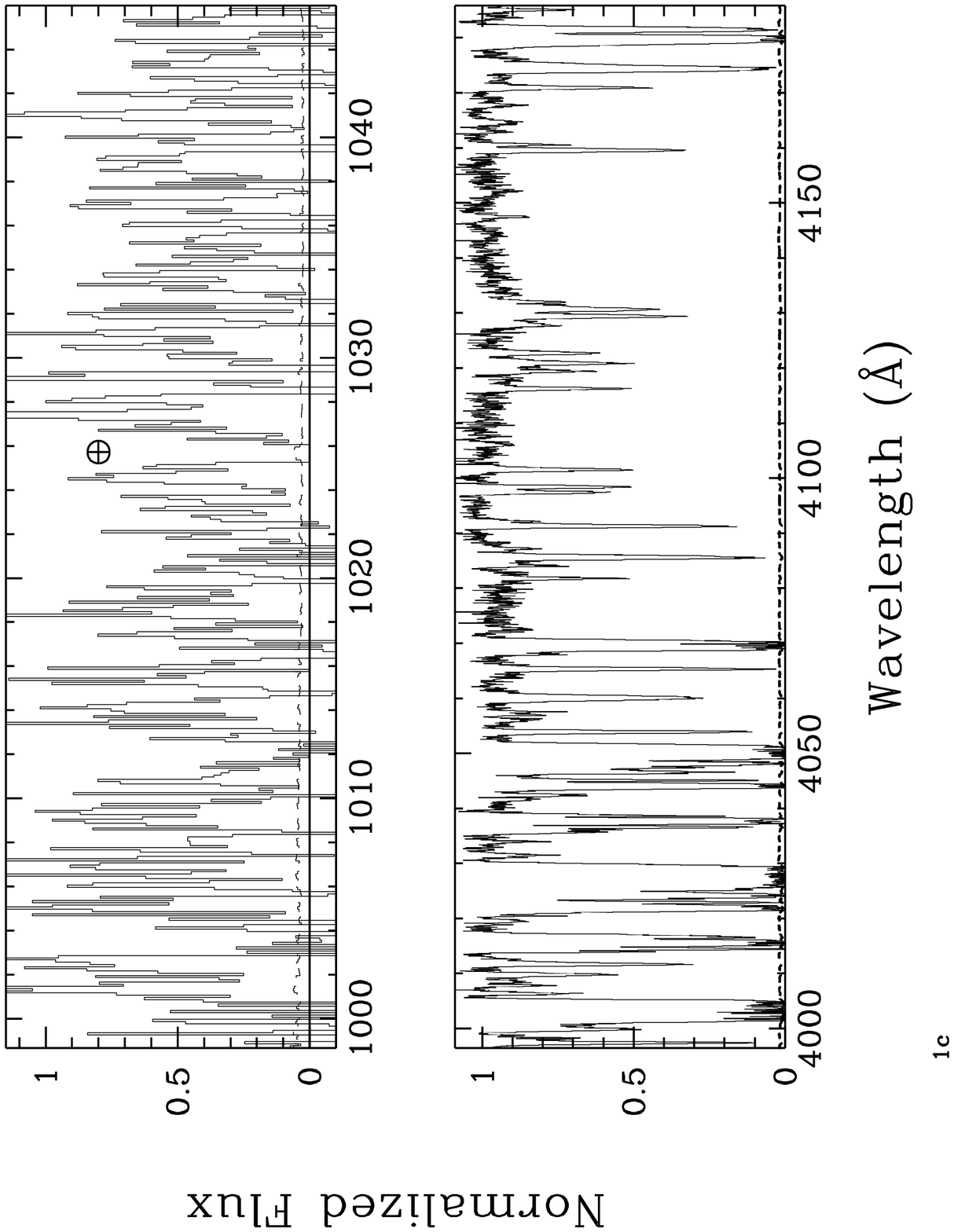}
\caption{c}
\end{figure}

\clearpage

\setcounter{figure}{0}
\begin{figure}
\plotone{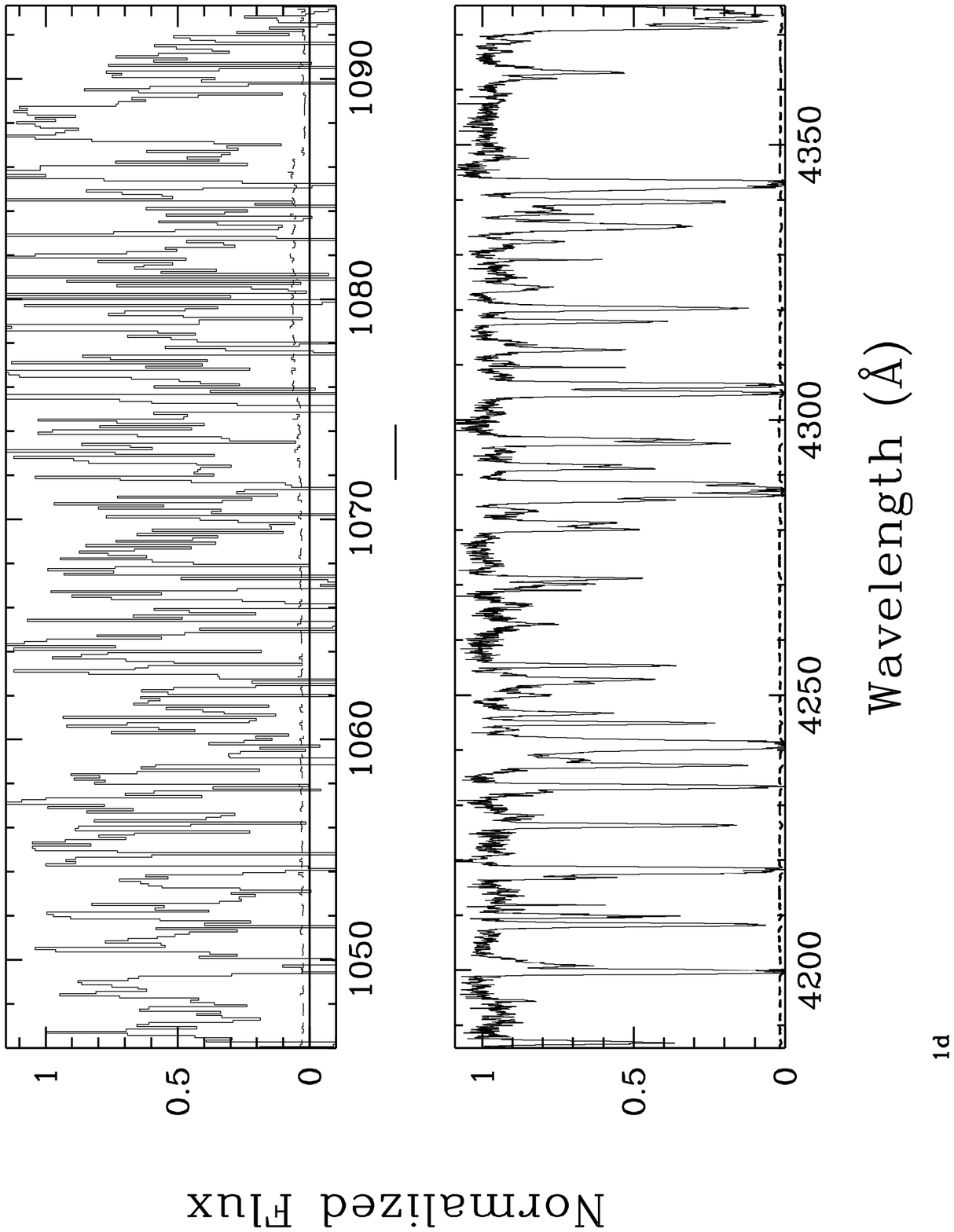}
\caption{d}
\end{figure}

\clearpage

\setcounter{figure}{0}
\begin{figure}
\plotone{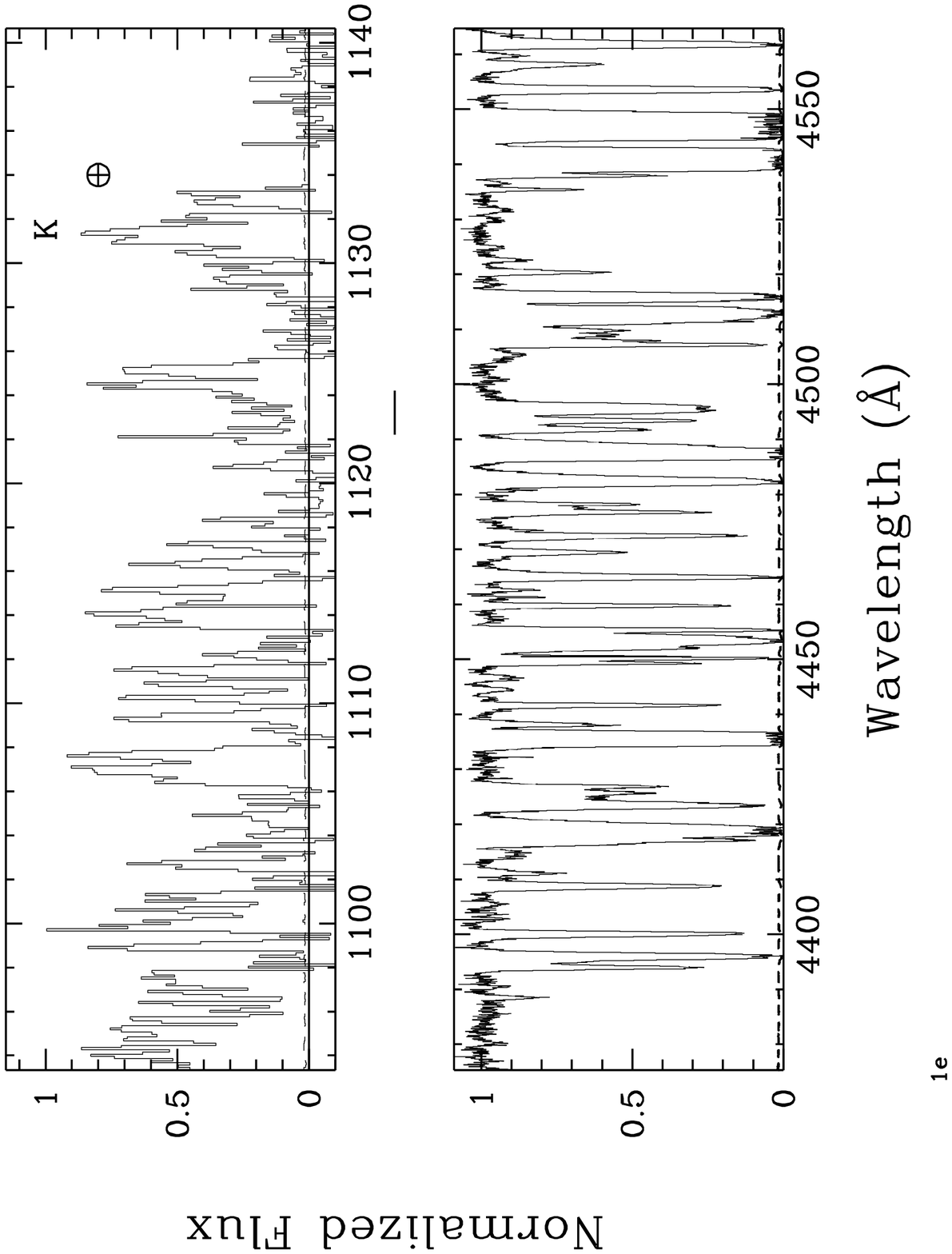}
\caption{e}
\end{figure}

\clearpage

\setcounter{figure}{0}
\begin{figure}
\plotone{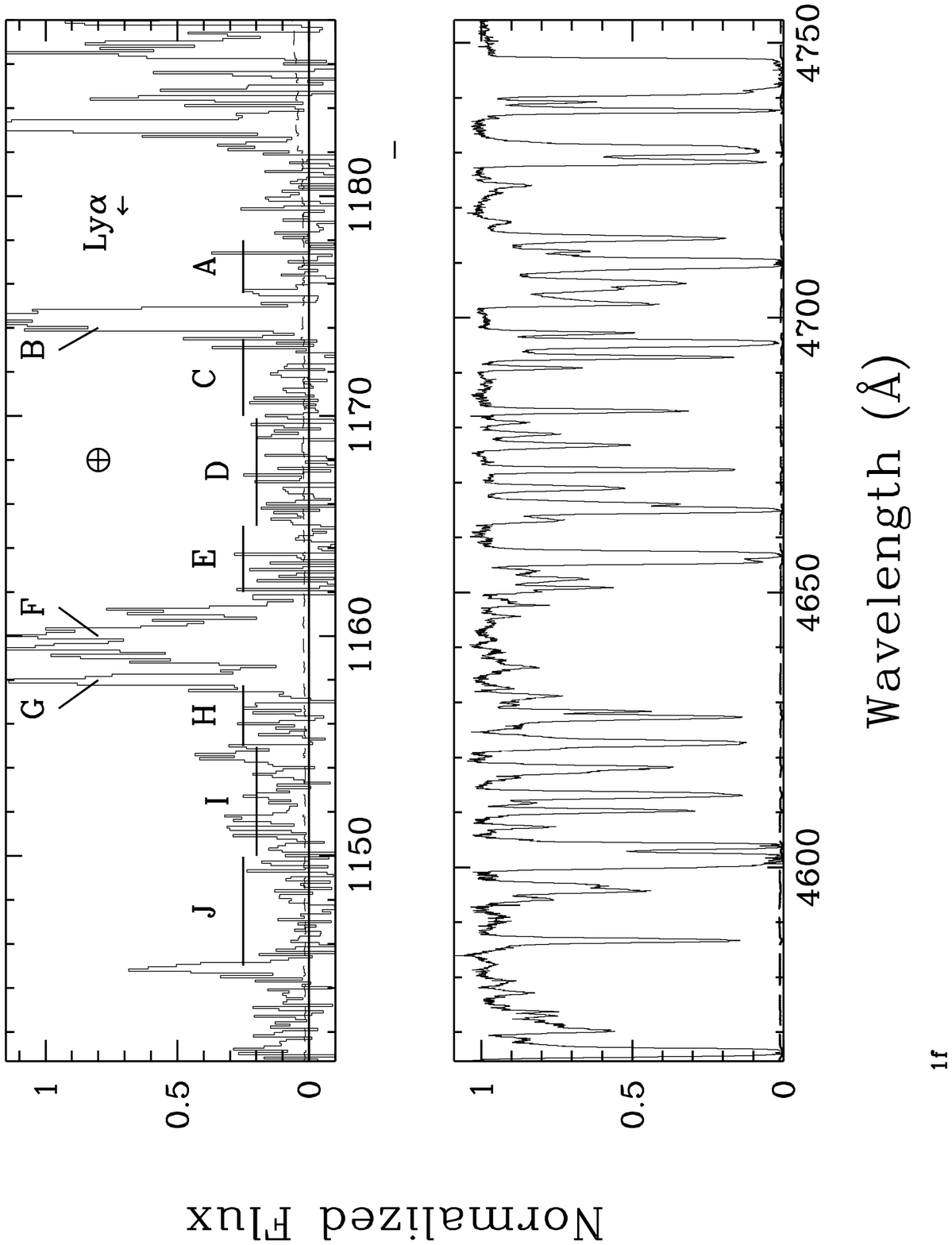}
\caption{f}
\end{figure}

\clearpage

\begin{figure}
\plotone{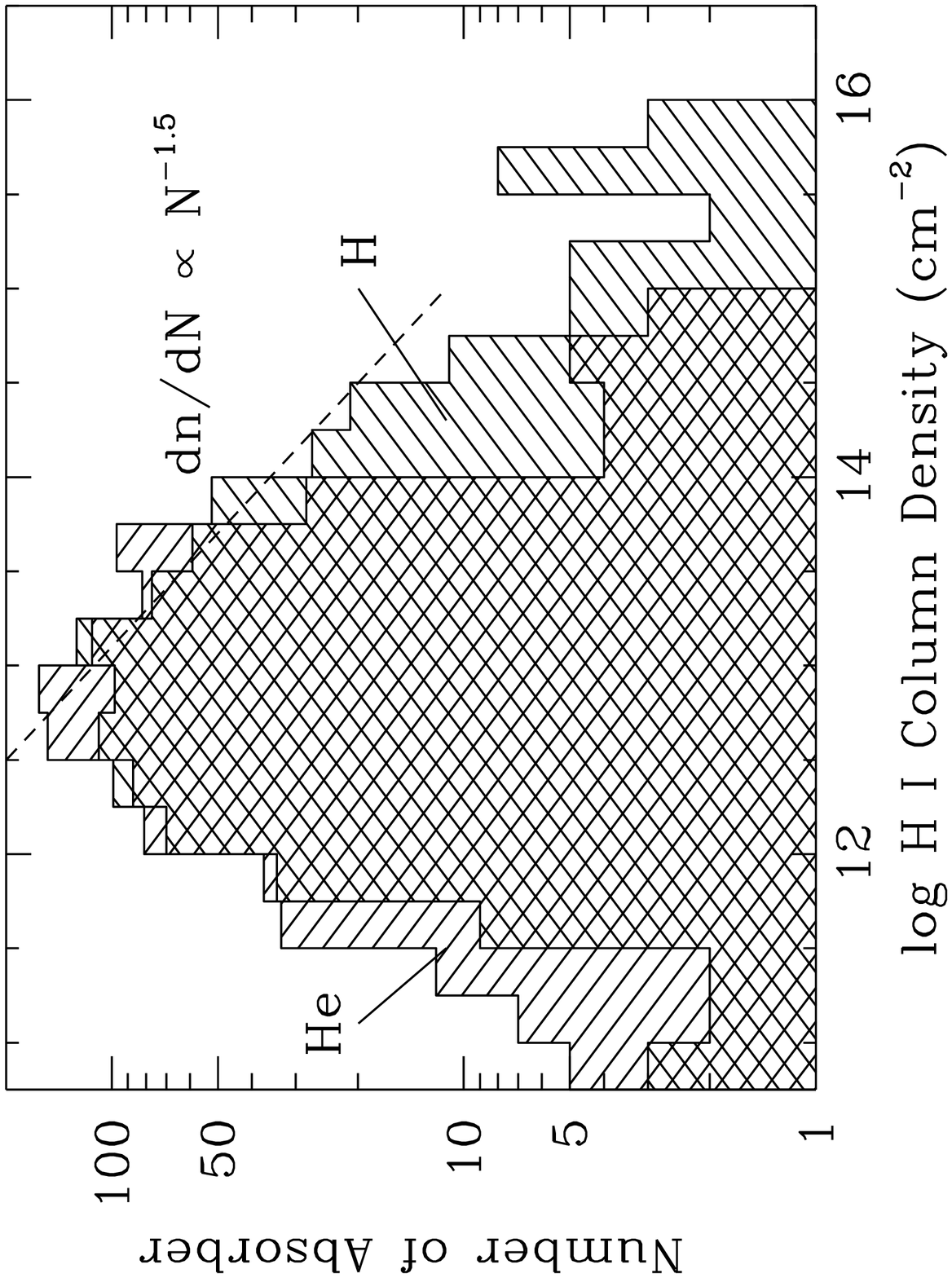}

\caption{~}
\end{figure}

\clearpage

\begin{figure}
\plotone{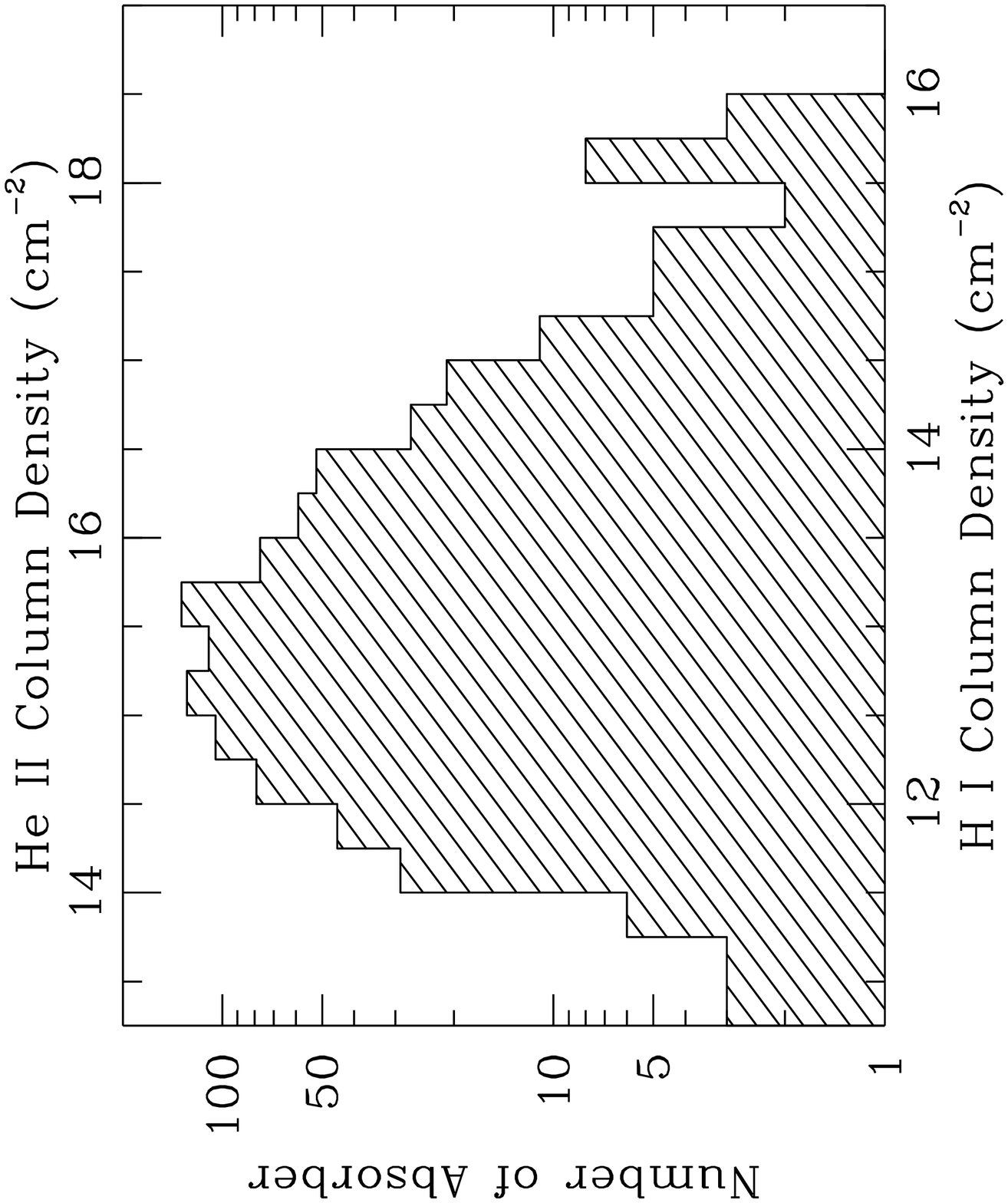}
\caption{~}
\end{figure}

\clearpage

\begin{figure}
\plotone{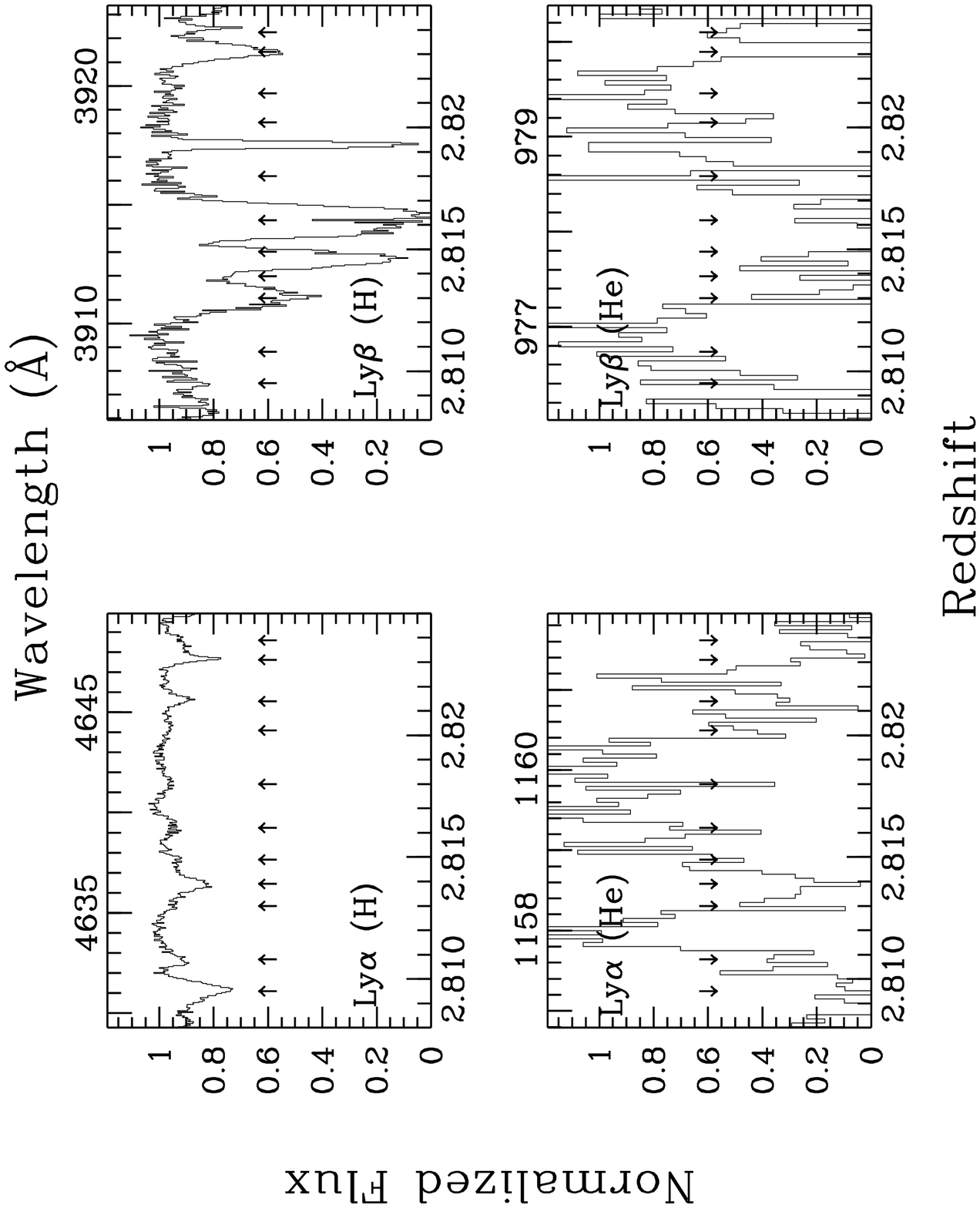}
\caption{~}
\end{figure}

\clearpage

\begin{figure}
\plotone{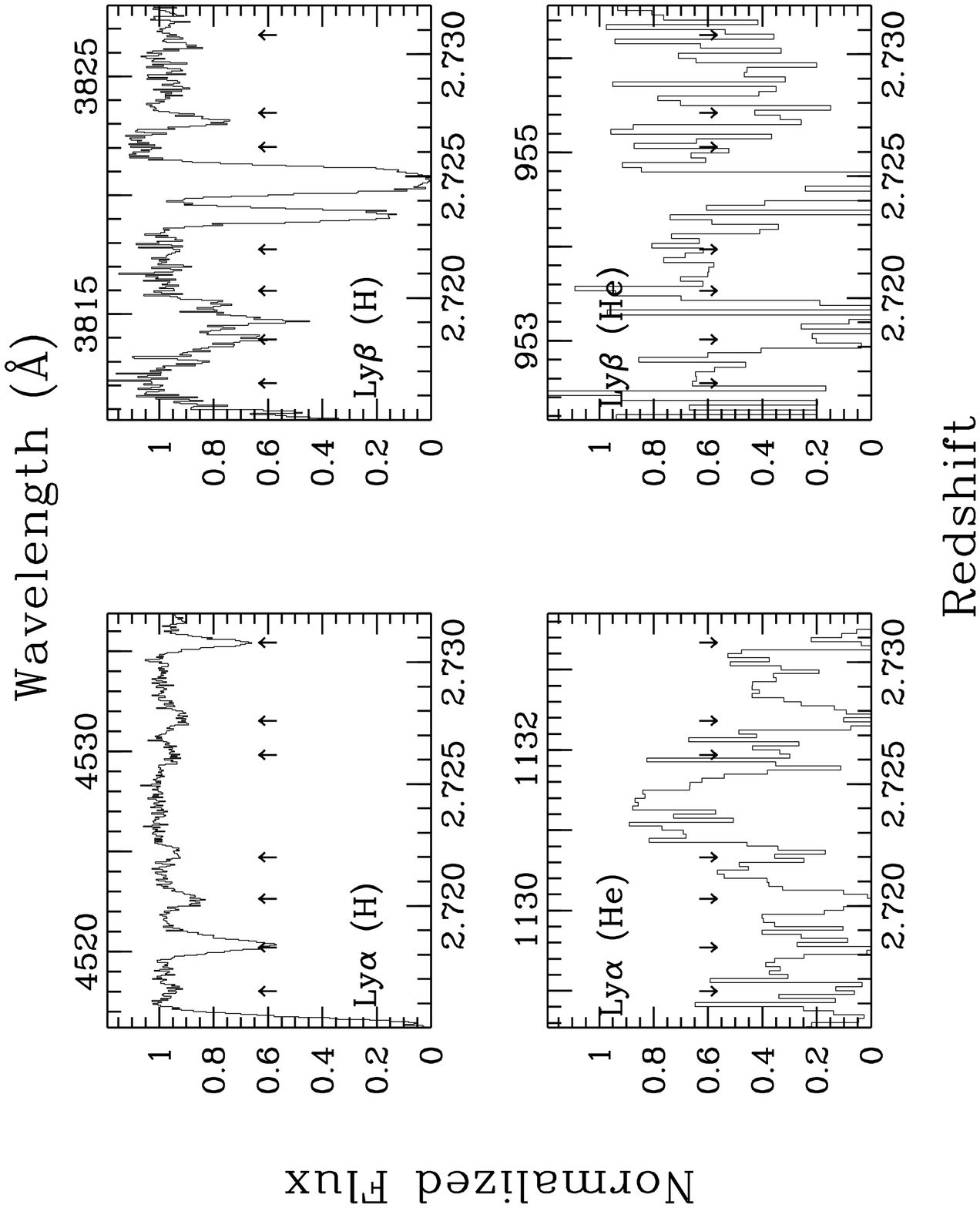}
\caption{~}
\end{figure}

\clearpage

\begin{figure}
\plotone{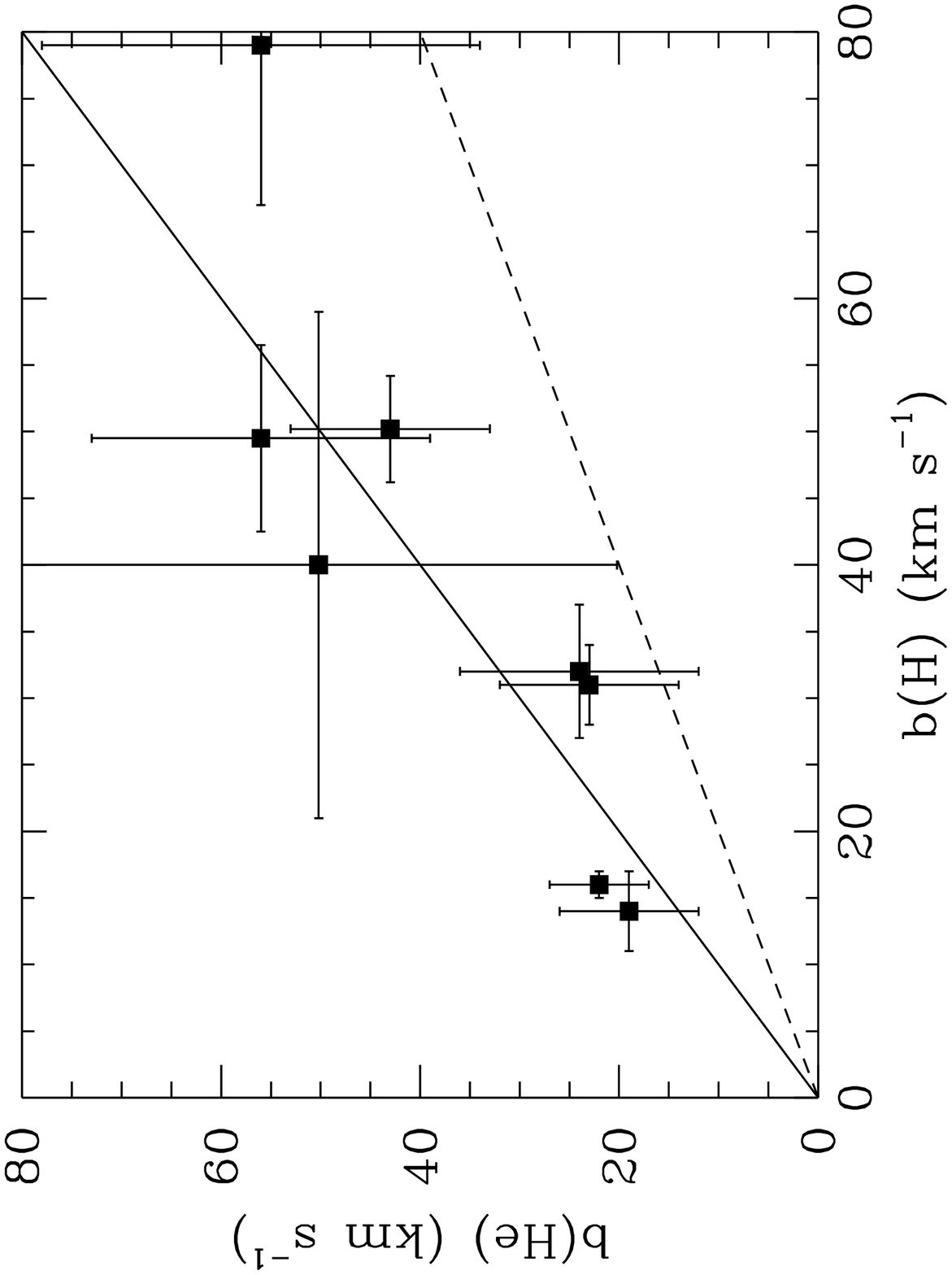}
\caption{~}
\end{figure}

\clearpage

\begin{figure}
\plotone{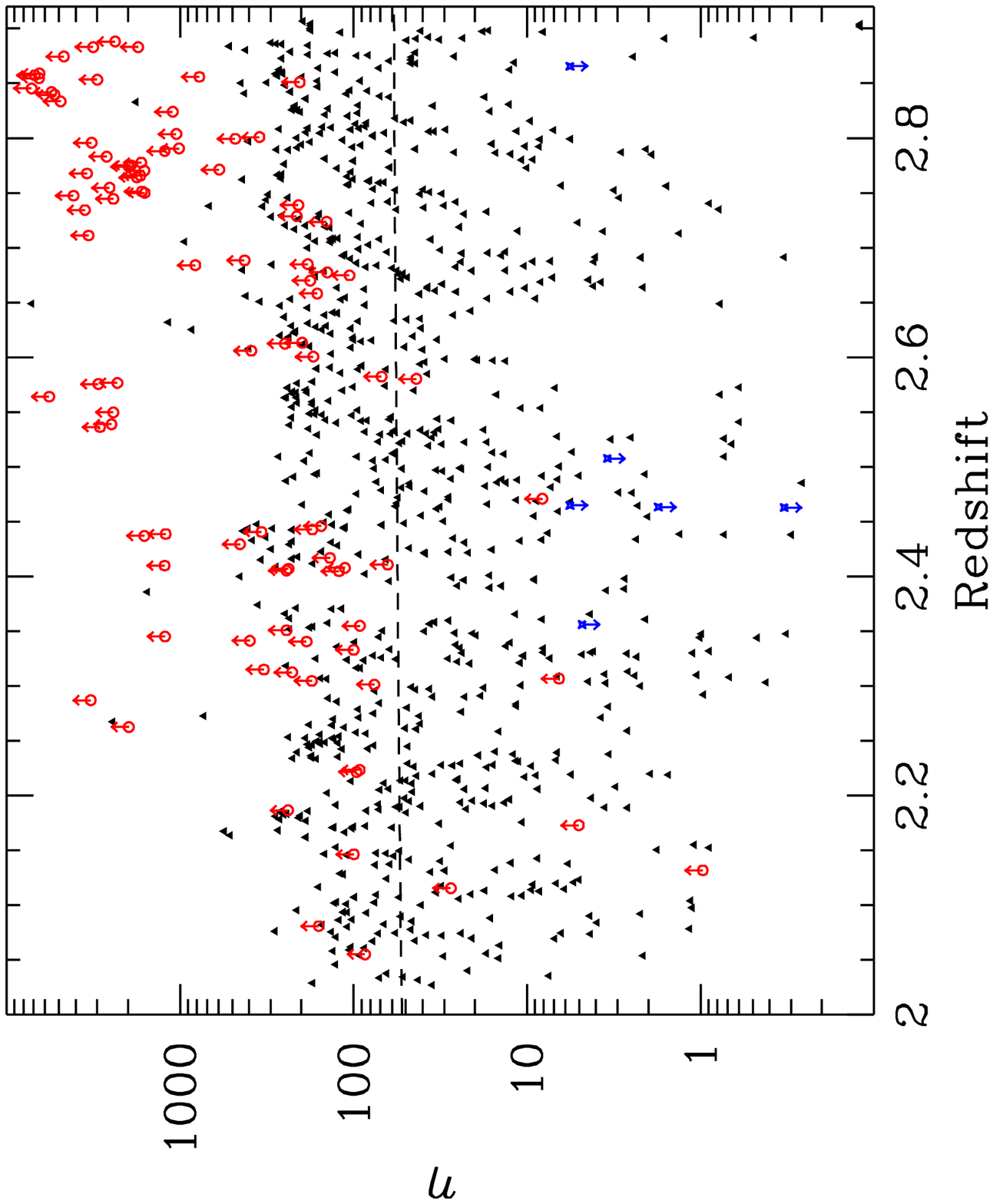}
\caption{~}
\end{figure}

\clearpage

\begin{figure}
\plotone{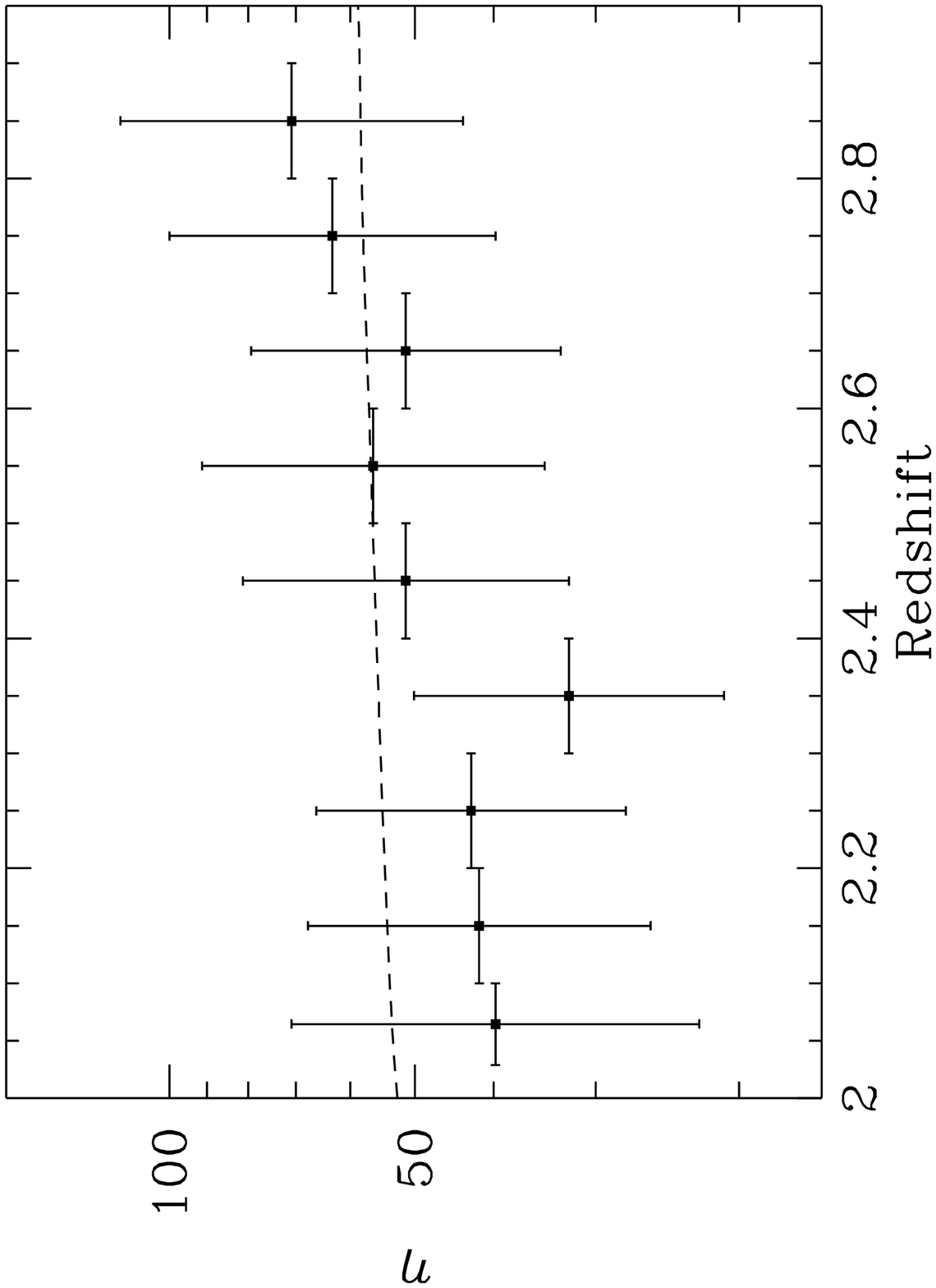}
\caption{~}
\end{figure}

\clearpage

\begin{figure}
\plotone{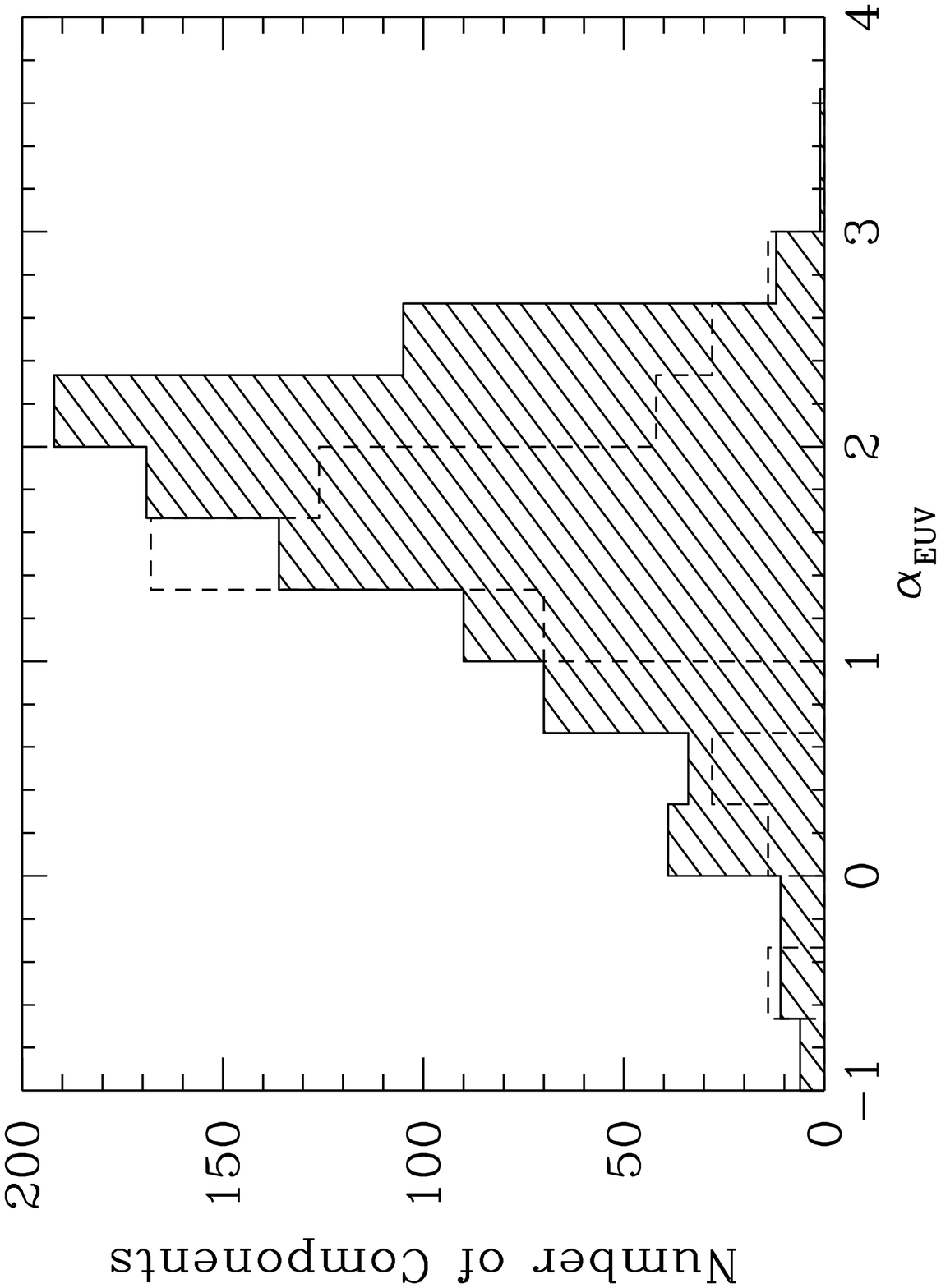}
\caption{~}
\end{figure}

\clearpage

\begin{figure}
\plotone{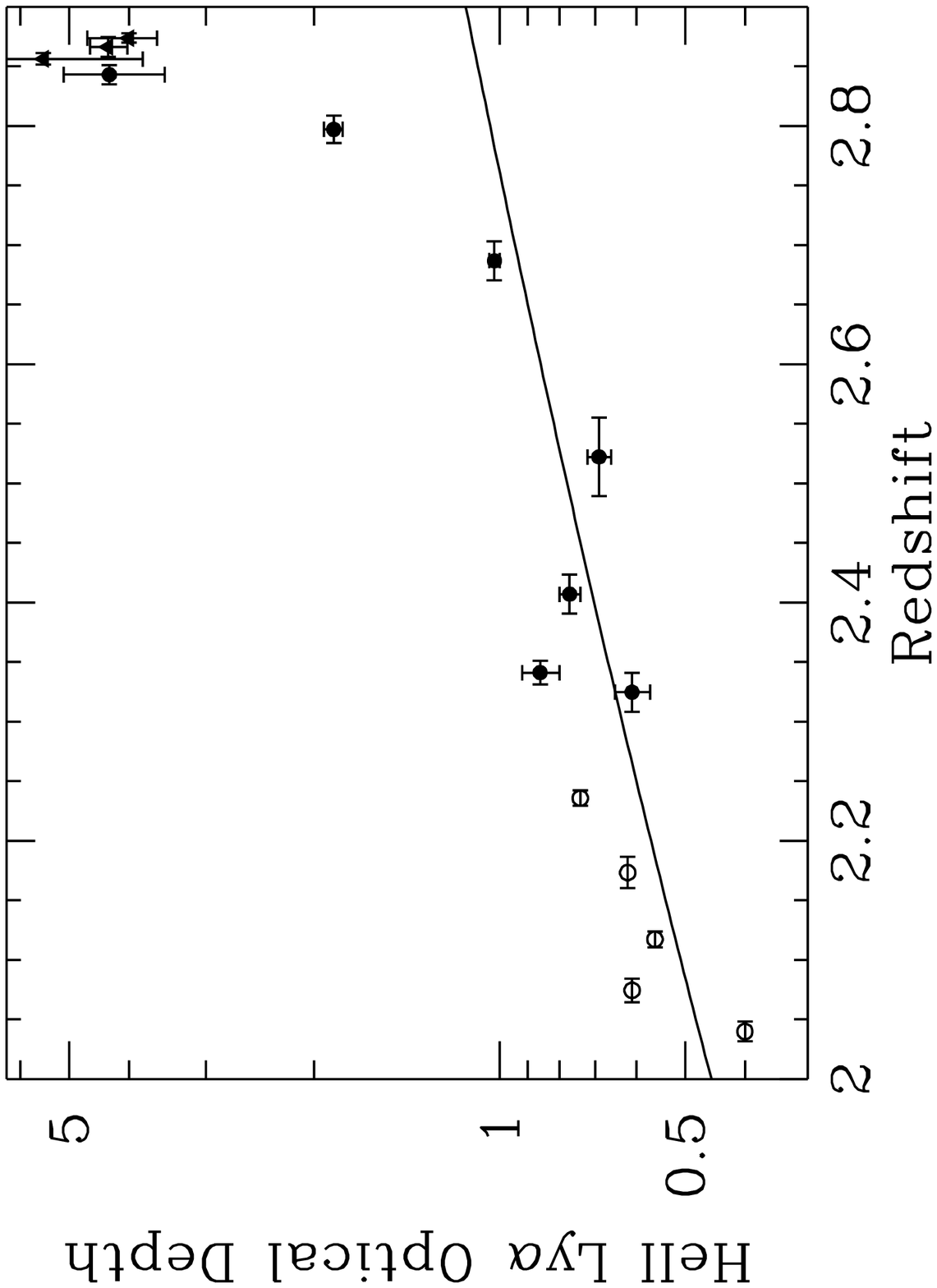}
\caption{~}
\end{figure}

\clearpage

{
\begin{deluxetable}{ccccc}
\tablecaption{Properties of Fitted Absorption Components\tablenotemark{a}
\label{tbl-1}}
\tablewidth{0pt}
\footnotesize
\tighttable
\tablehead{
\colhead{Redshift} &
\colhead{N(\ion{He}{2})} & 
\colhead{b\tablenotemark{b}} &
\colhead{N(\ion{H}{1})} &
\colhead{$\eta$}\\
\colhead{}      &
\colhead{$10^{12}$ \cl}  &
\colhead{\kms} &
\colhead{$10^{12}$ \cl}
}
\startdata
$ 2.8881 \pm  0.0002 $ & $1173 \pm  228 $ & $ 27.0 \pm  0.0$ & $  0.496 \pm0.000 $ & 2367\\
$2.8872 \pm 0.0009 $&$  616   \pm     242 $&$    26.2 \pm   67.4  $&$    2.05 \pm    3.73 $&  300  \\
$2.8861 \pm 0.0001 $&$  914   \pm    1040 $&$    22.1 \pm  3.20   $&$    3.27 \pm    1.96 $&  280\\
$2.8858 \pm 0.0001 $&$  1019  \pm    1706 $&$    30.3 \pm    13.6 $&$    3.92 \pm    2.15 $&     260 \\ 
$2.8848 \pm 0.0001 $&$    477 \pm     958 $&$    30.4 \pm    18.6 $&$    2.61 \pm    1.40 $&     182 \\ 
$2.8839 \pm 0.0003 $&$    442 \pm    3595 $&$    24.7 \pm    46.1 $&$   0.849 \pm    1.31 $&     520 \\ 
$2.8831 \pm 0.0002 $&$   1592 \pm   17923 $&$    27.0 \pm   0.0 $&$   0.502 \pm   0.000 $&    3172 \\ 
$2.8805 \pm 0.0000 $&$   1175 \pm    1456 $&$    19.8 \pm    3.40 $&$    2.80 \pm   0.653 $&     420 \\ 
$2.8788 \pm 0.0001 $&$    259 \pm    5190 $&$    17.7 \pm    6.30 $&$    1.12 \pm   0.131 $&     231 \\ 
$2.8781 \pm 0.0000 $&$   2918 \pm    7125 $&$    32.3 \pm   0.500 $&$    66.3 \pm   0.784 $&    44.0 \\ 
\enddata
\tablenotetext{a}{The complete version of this table is in the electronic 
edition of the Journal.  The printed edition contains only a sample.
}
\tablenotetext{b}{Zero error designates an added component with fixed value}
\end{deluxetable}
}

\clearpage

{
\begin{deluxetable}{cccccc}
\tablecaption{Absorption Lines in Spectral Voids\label{tbl-2}}
\tablewidth{0pt}
\footnotesize
\tighttable
\tablehead{
\colhead{Redshift} & 
\colhead{N(\he)}&
\colhead{N(\h)}&
\colhead{b(\he)}& 
\colhead{b(\h)}& 
\colhead{$\eta$}\\
& \multicolumn{2}{c}{($10^{12}$ \cl)}& \multicolumn{2}{c}{(\kms)}&
}
\startdata
2.8230 & $374\pm 100 $ & $3.9 \pm 0.4$ & $22 \pm 5$ & $16 \pm 1$ & $95\pm 26$ \\ 
2.8214 & $ 220\pm 22$ & $ 1.5\pm 0.3$ & $ 19\pm 7$ & $ 14\pm 3$ & $146\pm 55$ \\ 
2.8180 & $ 112\pm 42$& $ 2.0\pm 0.3$ & $24\pm 12$ & $32 \pm 5$ & $57\pm 24$ \\ 
2.8161 & $ 117\pm $ 61& $2.5\pm 0.3$ & $23 \pm 9$ & $ 31\pm 3$ & $ 47\pm 25$ \\ 
2.7276 & $ 777\pm 616$ & $ 4.8\pm 1.1$  & $88  \pm 66$ & $50 \pm 9$ & $160\pm 132$ \\ 
2.7262 & $ 617\pm 219$ & $ 3.0\pm 0.7$  & $56 \pm 17$ & $50 \pm 7$ & $206\pm 87$ \\ 
2.7220 & $ 235\pm 258$ & $2.6 \pm 0.2$ & $50 \pm 30$ & $40 \pm 19$ & $90\pm 98$ \\ 
2.7203 & $ 812\pm 468$ & $4.3\pm 0.6$ & $56 \pm 22$ & $79 \pm 12$ & $189 \pm 112$ \\ 
2.7183 & $ 2948\pm 1007$ & $ 21\pm 0.2$ & $43\pm 10$ & $50 \pm 4$ & $ 137\pm 47$\\ 
\enddata
\end{deluxetable}
}

\clearpage
{
\begin{deluxetable}{cccccccc}
\rotate
\tablecaption{\he\ \lya\ Optical Depths at $z > 2.77$\label{tbl-3}}
\tablewidth{0pt}
\footnotesize
\tighttable
\tablehead{
\colhead{Interval} &
\colhead{\he\ \lya\ range} & 
\colhead{$z$} &
\colhead{$\tau$(\lya)}&
\colhead{$\tau$(\lya)}&
\colhead{$\tau_{\rm raw}$(\lyb)}&
\colhead{$\tau_{\rm cor}$(\lyb)}&
\colhead{$\tau$(\lya)}\\
\colhead{}      &
\colhead{(\AA)} &
\colhead{}      &
\colhead{\FUSE} &
\colhead{STIS}  &
\colhead{\FUSE} &
\colhead{}      &
\colhead{Calculated}\\
\colhead{(1)}  &
\colhead{(2)}  &
\colhead{(3)}  &
\colhead{(4)}  &
\colhead{(5)}  &
\colhead{(6)}  &
\colhead{(7)}  &
\colhead{(8)} \\
}
\startdata
A & 1175.60--1178.00 & 2.8738 & $2.21\pm 0.20$ & $4.00^{+0.67}_{-0.40}$   & $1.10 \pm 0.18$& $0.75 \pm 0.18$  & $4.7 \pm 1.1$\\
B & 1174.00--1174.90 & 2.8661 & $0.07\pm0.05$ & $0.07 \pm 0.04$          & $\ldots$ & $\ldots$ & $\ldots$ \\
C & 1170.00--1173.50 & 2.8572 & $2.09 \pm 0.15$ & $4.80^{+\infty}_{-0.80}$ & $1.54 \pm 0.24$ & $ 0.89\pm 0.28$ & $5.5 \pm 1.7$\\
D & 1165.00--1169.90 & 2.8431 & $2.18\pm 0.14$ & $>4.54$                  & $1.04 \pm 0.11$ & $ 0.70\pm 0.13$ & $4.3 \pm 0.8$ \\
E & 1162.00--1165.00 & 2.8300 & $2.33\pm 0.18$ & $2.86^{+0.39}_{-0.28}$   & $0.72 \pm 0.11$ & $ 0.37 \pm 0.14$ & $2.3\pm 0.9$\\
F & 1159.00--1161.00 & 2.8185 & $0.33 \pm 0.03$ & $0.44 \pm 0.06$          & $\ldots$ & $\ldots$ & $\ldots$ \\
G & 1157.75--1158.25 & 2.8119 & $0.13\pm 0.06$ & $0.24^{+0.12}_{-0.11}$   & $\ldots$ & $\ldots$  & $\ldots$ \\
H & 1155.00--1157.75 & 2.8066 & $1.97\pm 0.12$ & $2.34^{+0.40}_{-0.28}$   & $0.91\pm 0.10$ & $ 0.56\pm 0.13$ & $3.5 \pm 0.8$ \\
I & 1150.00--1154.95 & 2.7938 & $1.79\pm 0.07$ & $3.72^{+\infty}_{-0.81}$ & $1.03\pm 0.10$ & $ 0.68 \pm 0.13$ & $ 4.3\pm0.8 $ \\
J & 1145.00--1149.95 & 2.7774 & $2.32\pm 0.12$ & $2.57^{+0.72}_{-0.41}$   & $0.94\pm 0.07$ & $ 0.60\pm 0.10$ & $ 3.8\pm 0.6$ \\
\enddata
\end{deluxetable}
}

\end{document}